\documentclass{article}
\usepackage{amssymb}
\usepackage{amsmath}
\usepackage{amsfonts}
\usepackage{graphicx}

\newtheorem{theorem}{Theorem}

\newtheorem{corollary}[theorem]{Corollary}

\newtheorem{definition}[theorem]{Definition}

\newtheorem{proposition}[theorem]{Proposition}
\newtheorem{remark}[theorem]{Remark}

\begin{document}

\author{
Ricardo A. Mosna$^{1,2}$\thanks{E-mail: mosna@ifi.unicamp.br}\hspace{0.2cm} and
Waldyr A. Rodrigues, Jr.$^{2}$\thanks{E-mail: walrod@mpc.com.br or walrod@ime.unicamp.br}\\
{\small $^{1}$Institute of Physics Gleb Wataghin}\\
{\small UNICAMP CP 6165 13083-970}\\
{\small Campinas, SP, Brazil}\\
{\small $\hspace{-0.8cm}^{2}$Institute of Mathematics, Statistics and Scientific Computation}\\
{\small IMECC-UNICAMP CP 6065 13083-970}\\
{\small Campinas, SP, Brazil}
}

\title{The Bundles of Algebraic and Dirac-Hestenes Spinor Fields\thanks{To appear
in Journal of Mathematical Physics \textbf{45}(7), 2945-2966 (2004).}
}

\maketitle

\begin{abstract}
The main objective of this paper is to clarify the \textit{ontology} of
Dirac-Hestenes spinor fields (\emph{DHSF}) \ and its relationship with even
multivector fields, on a Riemann-Cartan spacetime (\textit{RCST)}
$\mathfrak{M}=$($M,g,\nabla,\tau_{g},\uparrow$) admitting a spin structure,
and to give a mathematically rigorous derivation of the so called
Dirac-Hestenes equation (\textit{DHE}) \ in the case where $\mathfrak{M}$ is a
Lorentzian spacetime (the general case when $\mathfrak{M}$ is a \textit{RCST
}will be discussed in another publication). To this aim we introduce the
Clifford bundle of multivector fields ($\mathcal{C\ell}(M,g)$) and the
\emph{left }($C\ell_{\mathrm{Spin}_{1,3}^{e}}^{l}(M)$)\emph{\ }and
\emph{right} ($\mathcal{C\ell}_{\mathrm{Spin}_{1,3}^{e}}^{r}(M)$)
spin-Clifford bundles on the spin manifold $(M,g)$. The relation between
\textit{left ideal} \textit{algebraic spinor fields (LIASF)} and
Dirac-Hestenes \emph{spinor} fields (both fields are sections of
$\mathcal{C\ell}_{\mathrm{Spin}_{1,3}^{e}}^{l}(M)$) is clarified. We study in
detail the theory of covariant derivatives of Clifford fields as well as that
of left and right spin-Clifford fields. A consistent Dirac equation for a
\textit{DHSF }$\mathbf{\Psi}\in\sec\mathcal{C\ell}_{\mathrm{Spin}_{1,3}^{e}%
}^{l}(M)$ (denoted \textit{DE}$\mathcal{C\ell}^{l})$ on a Lorentzian spacetime
is found. We also obtain a \textit{representation} of the \textit{DE}%
$\mathcal{C\ell}^{l}$ in the Clifford bundle $\mathcal{C\ell}(M,g)$. It is
such equation that we call the \textit{DHE} and it is satisfied \ by Clifford
fields $\mathit{\psi}_{\Xi}\in\sec\mathcal{C\ell}(M,g)$. This means that to
each \textit{DHSF }$\mathbf{\Psi}\in\sec$ $\mathcal{C\ell}_{\mathrm{Spin}%
_{1,3}^{e}}^{l}(M)$ and to each spin frame $\Xi\in\sec P_{\mathrm{Spin}%
_{1,3}^{e}}(M)$, there is a well-defined sum of even multivector fields
$\mathit{\psi}_{\Xi}\in\sec\mathcal{C\ell}(M,g)$ (\emph{EMFS}) associated with
$\Psi$. Such an \emph{EMFS} is called a \textit{representative} of the
\emph{DHSF} on the given spin frame. And, of course, such a \emph{EMFS} (the
representative of the \emph{DHSF}) is \emph{not} a spinor field. With this
crucial distinction between a \emph{DHSF }and its \textit{representatives} on
the Clifford bundle, we provide a consistent theory for the covariant
derivatives of Clifford \ and spinor fields of all kinds. We emphasize that
the \textit{DE}$\mathcal{C\ell}^{l}$ and the \textit{DHE}, although related,
are equations of different mathematical natures. We study also the local
Lorentz invariance and the electromagnetic gauge invariance and show that only
for the \textit{DHE }such transformations are of the same mathematical nature,
thus suggesting a possible link between them.

\end{abstract}


\section{Introduction}

The main objective of this paper is to clarify the \textit{ontology }of
Dirac-Hestenes spinor fields (\textit{DHSF})\footnote{For the genesis of these
objects we quote \cite{23H}.} on general Riemann-Cartan spacetimes
(\textit{RCST)} and to give a mathematically justified account of the
Dirac-Hestenes equation (\textit{DHE})\textit{ }on Lorentzian spacetimes,
subjects that have been a matter of many misunderstandings and controversies
(as discussed in \cite{50}). Recall that the flat spacetime \textit{DHE}
represents the state of an electron by a map $\mathbf{\Psi}$ with values in
the even part of the Clifford algebra $\mathbb{R}_{1,3}$. However, a covariant
formulation of the \textit{DHE} on a (possibly curved) Lorentzian spacetime
$M$ cannot promote $\mathbf{\Psi}$, in a canonical way, to a section of the
Clifford bundle $\mathcal{C}\ell(M,g)$ (whose objects transform as tensors and
therefore cannot describe spin-1/2 particles). In \cite{50}, \textit{DHSF} on
a Minkowski spacetime were defined as equivalence classes of Clifford fields.
Here we follow a different approach, and define \textit{DHSF} as even sections
of an appropriate spinorial Clifford bundle. The objects satisfying the
Dirac-Hestenes equation are then even multivector fields which are
\emph{representatives} of \textit{DHSF} on the tensorial Clifford bundle.
Moreover, such a representative is manifestly spin-frame dependent, so that no
contradiction arises in representing spinors by Clifford fields.

To achieve our goals, we introduce in section 2 the Clifford bundle of
multivector fields\footnote{Of course, all the results of the present paper
could also be obtained in the case where $\mathcal{C\ell(}M,g)$ is a Clifford
bundle of nonhomogeneous differential forms.} ($\mathcal{C\ell}(M,g)$), and
the \emph{left (}$\mathcal{C\ell}_{\mathrm{Spin}_{1,3}^{e}}^{l}(M)$\emph{)
}and \emph{right} ($\mathcal{C\ell}_{\mathrm{Spin}_{1,3}^{e}}^{r}(M)$)
spin-Clifford bundles on the spin manifold $(M,g)$, and study in detail how
these bundles are related. Left algebraic spinor fields and Dirac-Hestenes
spinor fields (both fields are sections of $\mathcal{C\ell}_{\mathrm{Spin}%
_{1,3}^{e}}^{l}(M)$) are defined and the relation between them is established.
In section 4, a consistent Dirac equation for a \textit{DHSF }$\mathbf{\Psi
}\in\sec\mathcal{C\ell}_{\mathrm{Spin}_{1,3}^{e}}^{l}(M)$ (denoted
\textit{DE}$\mathcal{C\ell}^{l}$) on a Lorentzian manifold is found. In
section 5, we obtain a \textit{representation} of the \textit{DE}%
$\mathcal{C\ell}^{l}$ in the Clifford bundle, an equation we call the
Dirac-Hestenes equation (\textit{DHE}), which is satisfied\ by Clifford fields
$\mathit{\psi}_{\Xi}\in\sec\mathcal{C\ell}(M,g)$. This means that to each
\textit{DHSF }$\mathbf{\Psi}\in\sec$ $\mathcal{C\ell}_{\mathrm{Spin}_{1,3}%
^{e}}^{l}(M)$ and to each \textit{spin frame} $\Xi\in\sec P_{\mathrm{Spin}%
_{1,3}^{e}}(M)$ there is a well defined sum of even multivector fields
$\mathit{\psi}_{\Xi}\in\sec\mathcal{C\ell}(M,g)$ (\textit{EMFS}) associated
with $\Psi$. Such an \textit{EMFS} is called a \textit{representative} of the
\emph{DHSF} on the given spin frame. And, of course, such an \textit{EMFS}
(the representative of the \emph{DHSF}) is \emph{not} a spinor field. With
this crucial distinction between a \emph{DHSF} and their \textit{EMFS}
representatives, we present\ in section 5 an \textit{effective }spinorial
connection for the representatives of a \textit{DHSF} on $\mathcal{C\ell
}(M,g)$, thus providing a consistent theory for the covariant derivatives of
Clifford and spinor fields of all kinds.

We emphasize that the \textit{DE}$\mathcal{C\ell}^{l}$ and the \textit{DHE},
although related, are of different mathematical natures. This issue has been
\ particularly scrutinized in sections 4 and 5. We study also the local
Lorentz invariance and the electromagnetic gauge invariance and show that only
for the \textit{DHE }such transformations are of the same mathematical nature,
thus suggesting a possible link between them. In a sequel paper we are going
to investigate this issue and also (a) the formulation of the \textit{DE}%
$\mathcal{C\ell}$ and \textit{DHE} in an arbitrary Riemann-Cartan
spacetime through the use of a variational principle\footnote{We
shall use in our approach to the subject the techniques
of the multivector and extensor calculus developed in
\cite{16},\cite{17},\cite{18},\cite{45},\cite{46},\cite{47} and \cite{48}.};
(b) the theory of the Lie derivative of
the \textit{LIASF} and \textit{DHSF;} and (c) the claim in
\cite{22} that the existence of spinor fields in a Lorentzian
manifold requires a minimum amount of curvature. This problem is
important in view of the proposed teleparallel theories of the
gravitational field.

Finally, in the Appendix we derive some formulas employed in the main text for
the covariant derivative of Clifford and spinor fields, using the general
theory of covariant derivatives on associated vector bundles. In general, our
notation corresponds to that in \cite{50}.

A few acronyms are used in the present paper (to avoid long sentences) and
they are summarized below for the reader's convenience:

\textit{DHE}- Dirac-Hestenes Equation

\textit{DHSF- }Dirac-Hestenes Spinor Field

\textit{DE}$\mathcal{C\ell}^{l}$- Dirac equation for a \textit{DHSF
}$\mathbf{\Psi}\in\sec\mathcal{C\ell}_{\mathrm{Spin}_{1,3}^{e}}^{l}(M)$

\textit{EMFS- }Even Multivector Fields

\textit{LIASF- }Left Ideal Algebraic Spinor Field

\textit{PFB- }Principal Fiber Bundle

\textit{RIASF- }Right Ideal Algebraic Spinor Field

\textit{RCST- }Riemann-Cartan Spacetime

\section{The Clifford Bundle of Spacetime and their Irreducible Module
Representations}

\subsection{The Clifford Bundle of Spacetime}

Let $M$ be a four dimensional, real, connected, paracompact and noncompact
manifold. Let $TM$ [$T^{\ast}M$] be the tangent [cotangent] bundle of $M$.

\begin{definition}
A Lorentzian manifold is a pair $(M,g)$, where $g\in\sec T^{2,0}M$ is a
Lorentzian metric of signature $(1,3)$, i.e., for all $x\in M$, $T_{x}M\simeq
T_{x}^{*}M\simeq\mathbb{R}^{1,3}$, where $\mathbb{R}^{1,3}$ is the vector
Minkowski space.
\end{definition}

\begin{definition}
A spacetime $\mathfrak{M}$ is a pentuple $(M,g,\mathbf{\nabla,\tau}%
_{g},\mathbf{\uparrow})$ where $(M,g,\mathbf{\tau}_{g},\mathbf{\uparrow})$ is
an oriented Lorentzian manifold (oriented by $\mathbf{\tau}_{g}$) \ and time
oriented by an appropriated equivalence relation\footnote{See \cite{56} for
details.} (denoted $\uparrow$) for the timelike vectors at the tangent space
$T_{x}M$, $\forall x\in M$. $\mathbf{\nabla}$\textbf{\ }is a linear connection
for $M$ such that $\mathbf{\nabla}g=0$.
\end{definition}

\begin{definition}
Let $\mathbf{T}$ and $\mathbf{R}$ be respectively the torsion and curvature
tensors of $\nabla$. If in addition to the requirements of the previous
definitions, $\mathbf{T}(\mathbf{\nabla})=0$, then $\mathfrak{M}$ is said to
be a Lorentzian spacetime. The particular Lorentzian spacetime where
$M\simeq\mathbb{R}^{4}$ and such that $\mathbf{R}(\mathbf{\nabla})=0$ is
called Minkowski spacetime and will be denoted by $\mathcal{M}$. When
$\mathbf{T}(\mathbf{\nabla})$ is possibly nonzero, $\mathfrak{M}$ is said to
be a Riemann-Cartan spacetime (RCST). A particular RCST such that
$\mathbf{R}(\mathbf{\nabla})=0$ is called a teleparallel spacetime.
\end{definition}

In what follows $P_{\mathrm{SO}_{1,3}^{e}}(M)$ denotes the principal bundle of
oriented \textit{Lorentz tetrads}.\footnote{We assume that the reader is
acquainted with the structure of $P_{\mathrm{SO}_{1,3}^{e}}(M)$, whose
sections are the time oriented and oriented orthonormal frames, each one
\ associated by a local trivialization to a \textit{unique} element of
$SO_{1,3}^{e}(M).$ See, e.g., \cite{21},\cite{40},\cite{42} and \cite{43}.}

It is well known \cite{44Osborn} that the natural operations on metric vector
spaces, such as direct sum, tensor product, exterior power, etc., carry over
canonically to vector bundles with metrics.

\begin{definition}
The Clifford bundle of \ the Lorentzian manifold $(M,g)$ is the bundle of
algebras
\begin{equation}
\mathcal{C}\ell(M,g)=\bigcup_{x\in M}\mathcal{C}\ell(T_{x}M,g_{x}),
\label{1.1}%
\end{equation}
where $\mathcal{C}\ell(T_{x}M,g_{x})$ is the Clifford algebra associated with
$(T_{x}M,g_{x})$ (see, e.g., \cite{50}).
\end{definition}

As is well known (\cite{6},\cite{7},\cite{14}) $\mathcal{C}\ell(M,g)$ is a
\emph{quotient} (or \emph{factor}) bundle, namely
\begin{equation}
\mathcal{C}\ell(M,g)=\frac{\tau M}{\mathcal{J}(M,g)}\label{1.1bis}%
\end{equation}
where $\tau M={\oplus_{r=0}^{\infty}}T^{0,r}M$ and $T^{(0,r)}M$ is the space
of $r$-contravariant tensor fields, and $\mathcal{J}(M,g)$ is the bundle of
ideals whose fibers are the two-sided ideals in $\tau M$ generated by the
elements of the form $a\otimes b+b\otimes a-2g(a,b),$ with $a,b\in TM$. In
what follows, we denote the real Clifford algebra associated with $\mathbb{R}%
^{p,q}$ by $\mathbb{R}_{p,q}$. The even subalgebra of $\mathbb{R}_{p,q}$ will
be denoted by $\mathbb{R}_{p,q}^{0}$ (see, e.g., \cite{50}).

Let $\mathbf{\pi}_{c}:\mathcal{C}\ell(M,g)\rightarrow M$ be the canonical
projection of $\ \mathcal{C}\ell(M,g)$ and let $\{U_{\alpha}\}$ be an open
covering of $M$. There are trivialization mappings $\mathbf{\psi}%
_{i}:\mathbf{\pi}_{c}^{-1}(U_{i})\rightarrow U_{i}\times\mathbb{R}_{1,3}$ of
the form $\mathbf{\psi}_{i}(p)=(\mathbf{\pi}_{c}(p),\psi_{i,x}(p))=(x,\psi
_{i,x}(p))$. If $x\in U_{i}\cap U_{j}$ and $p\in\mathbf{\pi}_{c}^{-1}(x)$,
then
\begin{equation}
\psi_{i,x}(p)=h_{ij}(x)\psi_{j,x}(p)
\end{equation}
for $h_{ij}(x)\in\mathrm{Aut}(\mathbb{R}_{1,3})$, where $h_{ij}:U_{i}\cap
U_{j}\rightarrow\mathrm{Aut}(\mathbb{R}_{1,3})$ are the transition mappings of
$\mathcal{C}\ell(M,g)$. We know that every automorphism of $\mathbb{R}_{1,3}$
is \textit{inner} and it follows that
\begin{equation}
h_{ij}(x)\psi_{j,x}(p)=g_{ij}(x)\psi_{i,x}(p)g_{ij}(x)^{-1} \label{1.4}%
\end{equation}
for some $g_{ij}(x)\in\mathbb{R}_{1,3}^{\star}$, the group of invertible
elements of $\mathbb{R}_{1,3}$.

Now, the group $\mathrm{SO}_{1,3}^{e}$ has as it is well known (see, e.g.,
\cite{4},\cite{5},\cite{7},\cite{39},\cite{50}) a natural extension in the
Clifford algebra $\mathbb{R}_{1,3}$. Indeed we know that $\mathbb{R}%
_{1,3}^{\star}$ acts naturally on $\mathbb{R}_{1,3}$ as an algebra
automorphism through its adjoint representation. A set of \emph{lifts} of the
transition functions of $\mathcal{C}\ell(M,g)$ is a set of $\mathbb{R}%
_{1,3}^{\star}$-valued functions $\{g_{ij}\}$ such that if%
\begin{align}
\mathrm{Ad} &  :g\mapsto\mathrm{Ad}_{g},\nonumber\\
\mathrm{Ad}_{g}(a) &  =gag^{-1},\forall a\in\mathbb{R}_{1,3},\label{1.6bis}%
\end{align}
then $\mathrm{Ad}_{g_{ij}}=h_{ij}$ in all intersections.

Also\footnote{Recall that $\mathrm{Spin}_{1,3}^{e}=\{a\in\mathbb{R}_{1,3}%
^{0}:a\tilde{a}=1\}\simeq SL(2,\mathbb{C)}$ is the universal covering group of
the restricted Lorentz group $\mathrm{SO}_{1,3}^{e}$. See, e.g., \cite{50}.}
$\sigma=\mathrm{Ad}|_{\mathrm{Spin}_{1,3}^{e}}$ defines a group homomorphism
$\sigma:\mathrm{Spin}_{1,3}^{e}\rightarrow\mathrm{SO}_{1,3}^{e}$ which is onto
with kernel $\mathbb{Z}_{2}$. We have that Ad$_{-1}=$ identity, and so
$\mathrm{Ad}:\mathrm{Spin}_{1,3}^{e}\rightarrow\mathrm{Aut}(\mathbb{R}_{1,3})$
descends to a representation of $\mathrm{SO}_{1,3}^{e}$. Let us call
$\mathrm{Ad}^{\prime}$ this representation, i.e., $\mathrm{Ad}^{\prime
}:\mathrm{SO}_{1,3}^{e}\rightarrow\mathrm{Aut}(\mathbb{R}_{1,3})$. Then we can
write $\mathrm{Ad}_{\sigma(g)}^{\prime}a=\mathrm{Ad}_{g}a=gag^{-1}$.

From this it is clear that the structure group of the Clifford bundle
$\mathcal{C}\ell(M,g)$ is reducible from $\mathrm{Aut}(\mathbb{R}_{1,3})$ to
$\mathrm{SO}_{1,3}^{e}$. This follows immediately from the Lorentzian
structure of $(M,g)$ and the fact that $\mathcal{C}\ell(M,g)$ is the exterior
bundle where the fibers are equipped with the Clifford product. Thus the
transition maps of the principal bundle of oriented Lorentz tetrads
$P_{\mathrm{SO}_{1,3}^{e}}(M)$ can be (through $\mathrm{Ad}^{\prime}$) taken
as transition maps for the Clifford bundle. We then have \cite{7}
\begin{equation}
\mathcal{C}\ell(M,g)=P_{\mathrm{SO}_{1,3}^{e}}(M)\times_{\mathrm{Ad}^{\prime}%
}\mathbb{R}_{1,3},
\end{equation}
i.e., the Clifford bundle is an associated vector bundle to the principal
bundle $P_{\mathrm{SO}_{1,3}^{e}}(M)$ of orthonormal Lorentz frames.

\begin{definition}
Sections of $\mathcal{C}\ell(M,g)$ are called Clifford fields.\footnote{We
note that the term Clifford fields was used in \cite{50} for mappings from
Minkowski spacetime to the Clifford algebra $\mathbb{R}_{1,3}$.}
\end{definition}

\subsection{Spinor Bundles}

\begin{definition}
\label{spin structure}A spin structure on $M$ consists of a principal fiber
bundle $\mathbf{\pi}_{s}:P_{\mathrm{Spin}_{1,3}^{e}}(M)\rightarrow M$
\ (called the Spin Frame Bundle), with group $\mathrm{Spin}_{1,3}^{e}$, and a
map
\begin{equation}
s:P_{\mathrm{Spin}_{1,3}^{e}}(M)\rightarrow P_{\mathrm{SO}_{1,3}^{e}%
}(M)\label{spinor bundle 1}%
\end{equation}
satisfying the following conditions
\end{definition}

(i) $\mathbf{\pi}(s(p))=\mathbf{\pi}_{s}(p)\ \forall p\in P_{\mathrm{Spin}%
_{1,3}^{e}}(M);$ $\pi$ is the projection map of the bundle $P_{\mathrm{SO}%
_{1,3}^{e}}(M)$.

(ii) $s(pu)=s(p)Ad_{u}\ ,\forall p\in P_{\mathrm{Spin}_{1,3}^{e}}(M)$ and
$Ad:\mathrm{Spin}_{1,3}^{e}\rightarrow\mathrm{Aut}(\mathbb{R}_{1,3}),$
$Ad_{u}:\mathbb{R}_{1,3}\ni x\mapsto uxu^{-1}\in\mathbb{R}_{1,3}$.\medskip

Recall that minimal left (right) ideals of $\mathbb{R}_{p,q}$ are left (right)
modules for $\mathbb{R}_{p,q}$ \cite{50}. In \cite{50}, covariant, algebraic
and Dirac-Hestenes spinors (when ($p,q)=(1,3)$) were defined as certain
equivalence classes in appropriate sets, and a \emph{preliminary} definition
for fields of these objects living on \textit{Minkowski} spacetime was given.
We are now interested in defining algebraic Dirac spinor fields and also
Dirac-Hestenes spinor fields, on a general Riemann-Cartan spacetime
(definition 3), as sections of appropriate vector bundles (spinor bundles)
associated to $P_{\mathrm{Spin}_{1,3}^{e}}(M)$. The compatibility between
$P_{\mathrm{Spin}_{1,3}^{e}}(M)$ and $P_{\mathrm{SO}_{1,3}^{e}}(M)$, as
captured in definition \ref{spin structure}, is essential for that matter.

It is therefore natural to ask the following: When does a spin structure exist
on an oriented manifold $M$? The answer, which is a classical result
(\cite{1},\cite{6},\cite{7},\cite{14},\cite{20},\cite{40},\cite{42}%
-\cite{44},\cite{44p},\cite{44Osborn}), is that the necessary and sufficient
conditions for the existence of a spin structure on $M$ is that the second
Stiefel-Whitney class $w_{2}(M)$ of $M$ is trivial. Moreover, when a spin
structure exists, one can show that it is unique (modulo isomorfisms) if and
only if $H^{1}(M,\mathbb{Z}_{2})$ is trivial.

\begin{remark}
\label{geroch rem 1}For a spacetime $\mathfrak{M}$ (definition 2), a spin
structure exists if and only if $P_{\mathrm{SO}_{1,3}^{e}}(M)$ is a
\emph{trivial} bundle. This was originally shown by Geroch \cite{21}.
\end{remark}

\begin{definition}
\label{SPIN FRAME}We call global sections $\xi\in\sec P_{\mathrm{SO}_{1,3}%
^{e}}(M)$ \emph{Lorentz frames }and global sections $\Xi\in\sec
P_{\mathrm{Spin}_{1,3}^{e}}(M)$ \emph{spin frames}.
\end{definition}

\begin{remark}
\label{geroch rem 2}Recall that a principal bundle is trivial if and only if
it admits a global section. Therefore, Geroch's result says that\ a
(noncompact) spacetime admits a spin structure if and only if it admits a
(globally defined) Lorentz frame. In fact, it is possible to replace
$P_{\mathrm{SO}_{1,3}^{e}}(M)$ by $P_{\mathrm{Spin}_{1,3}^{e}}(M)$ in remark
\ref{geroch rem 1} (see \cite{21}, footnote 25). In this way, when a
(noncompact) spacetime admits a spin structure, the bundle $P_{\mathrm{Spin}%
_{1,3}^{e}}(M)$ is trivial and, therefore, every bundle associated to it is
also trivial.
\end{remark}

\begin{definition}
An oriented manifold endowed with a spin structure will be called a spin manifold.
\end{definition}

We now present the most usual definitions of spinor bundles appearing in the
literature\footnote{We recall that there are some other (equivalent)
definitions of spinor bundles that we are not going to introduce in this paper
as, e.g., the one given in \cite{7bleecker} in terms of mappings from
\ $P_{\mathrm{Spin}_{1,3}^{e}}$ to some appropriate vector space.} and next we
find appropriate vector bundles such that particular sections are
\textit{LIASF} or \textit{DHSF}.

\begin{definition}
\label{LRSB}A real spinor bundle for $M$ is a vector bundle
\begin{equation}
S(M)=P_{\mathrm{Spin}_{1,3}^{e}}(M)\times_{\mu_{l}}\mathbf{M} \label{1.7}%
\end{equation}
where $\mathbf{M}$ is a left module for $\mathbb{R}_{1,3}$ and $\mu_{l}$ is a
representation of $\mathrm{Spin}_{1,3}^{e}$ on $End(\mathbf{M)}$ given by left
multiplication by elements of $\mathrm{Spin}_{1,3}^{e}$.
\end{definition}

\begin{definition}
The dual bundle $S^{\star}(M)$ is a real spinor bundle
\begin{equation}
S^{\star}(M)=P_{\mathrm{Spin}_{1,3}^{e}}(M)\times_{\mu_{r}}\mathbf{M}^{\star
}\label{1.7bis}%
\end{equation}
where $\mathbf{M}^{\star}$ is a right module for $\mathbb{R}_{1,3}$ and
$\mu_{r}$ is a representation of $\mathrm{Spin}_{1,3}^{e}$ in \textrm{End}%
$(\mathbf{M)}$ given by right multiplication\footnote{More precisely, this
means that given $u\in\mathrm{Spin}_{1,3}^{e},$ $a\in\mathbf{M}^{\star},$
$\mu_{r}(u)a=au^{-1},$ so that $\mu_{r}(uu')a=a(uu')^{-1}=au'^{-1}u^{-1}=
\mu_{r}(u)\mu_{r}(u')a$.\label{fnr}} by (inverse) elements of
$\mathrm{Spin}_{1,3}^{e}$.
\end{definition}

\begin{definition}
A complex spinor bundle for $M$ is a vector bundle
\begin{equation}
S_{c}(M)=P_{\mathrm{Spin}_{1,3}^{e}}(M)\times_{\mu_{c}}\mathbf{M}_{c}%
\end{equation}
where $\mathbf{M}_{c}$ is a complex left module for $\mathbb{C}\otimes
\mathbb{R}_{1,3}\simeq\mathbb{R}_{4,1}\simeq\mathbb{C(}4\mathbb{)}$, and where
$\mu_{c}$ is a representation $\ $of $\mathrm{Spin}_{1,3}^{e}$ in
$\mathrm{End}(\mathbf{M}_{c})$ given by left multiplication by elements of
$\mathrm{Spin}_{1,3}^{e}$.
\end{definition}

\begin{definition}
The dual complex spinor bundle for $M$ is a vector bundle
\begin{equation}
S_{c}^{\star}(M)=P_{\mathrm{Spin}_{1,3}^{e}}(M)\times_{\mu_{c}}\mathbf{M}%
_{c}^{\star}%
\end{equation}
where $\mathbf{M}_{c}^{\star}$ is a complex right module for $\mathbb{C}%
\otimes\mathbb{R}_{1,3}\simeq\mathbb{R}_{4,1}\simeq\mathbb{C(}4\mathbb{)}$,
and where $\mu_{c}$ is a representation $\ $of $\mathrm{Spin}_{1,3}^{e}$ in
\textrm{End}$(\mathbf{M}_{c})$ given by right multiplication$^{\ref{fnr}}$ by
(inverse) elements of $\mathrm{Spin}_{1,3}^{e}$.
\end{definition}

Taking, e.g., $\mathbf{M}_{c}=\mathbb{C}^{4}$ and $\mu_{c}$ the $D^{(1/2,0)}%
\oplus D^{(0,1/2)}$ representation of $\mathrm{Spin}_{1,3}^{e}\cong
SL(2,\mathbb{C})$ in $\mathrm{End}(\mathbb{C}^{4})$, we immediately recognize
the usual definition of the covariant spinor bundle of $M$ as given, e.g., in
\cite{11},\cite{12},\cite{13},\cite{20},\cite{42} and \cite{43}.

\subsection{Left Spin-Clifford Bundle}

As shown in \cite{50}, besides the ideal $I=\mathbb{R}_{1,3}\frac{1}%
{2}(1+E_{0})$, other ideals exist in $\mathbb{R}_{1,3}$ that are only
\textit{algebraically} equivalent to this one. (This fact gives rise
to a large class of multivector Dirac equations in flat spacetime,
generalizing the Dirac-Hestenes equation \cite{41mosna1,41MOSNA2}.) In order
to capture all possibilities we recall that $\mathbb{R}_{1,3}$ can be
considered as a module over itself by left (or right) multiplication. We are
thus led to the following definition.

\begin{definition}
\label{LsCbundle}The left real spin-Clifford bundle of $M$ is the vector
bundle
\begin{equation}
\mathcal{C}\ell_{\mathrm{Spin}_{1,3}^{e}}^{l}(M)=P_{\mathrm{Spin}_{1,3}^{e}%
}(M)\times_{l}\mathbb{R}_{1,3}%
\end{equation}
where $l$ is the representation of $\mathrm{Spin}_{1,3}^{e}$ on $\mathbb{R}%
_{1,3}$ given by $l(a)x=ax$. Sections of $\mathcal{C}\ell_{\mathrm{Spin}%
_{1,3}^{e}}^{l}(M)$ are called left spin-Clifford fields.
\end{definition}

\begin{remark}
\label{''principalbunde''}$\mathcal{C}\ell_{\mathrm{Spin}_{1,3}^{e}}^{l}(M)$
is a \textquotedblleft principal $\mathbb{R}_{1,3}$- bundle\textquotedblright,
i.e., it admits a free action of $\mathbb{R}_{1,3}$ on the right \cite{7},
which is denoted by $R_{g}$, $g\in\mathbb{R}_{1,3}$. This will be considered
in section 5.
\end{remark}

\begin{remark}
There is a \emph{natural} embedding\footnote{The symbol $A\hookrightarrow B$
means that $A$ is embedded in $B$ and $A\subseteq B$.} $P_{\mathrm{Spin}%
_{1,3}^{e}}(M)\hookrightarrow\mathcal{C}\ell_{\mathrm{Spin}_{1,3}^{e}}^{l}(M)$
which comes from the embedding $\mathrm{Spin}_{1,3}^{e}\hookrightarrow
\mathbb{R}_{1,3}^{0}$. Hence (as we shall see in more details below), every
real left spinor bundle (definition \ref{LsCbundle}) for $M$ can be
captured from $\mathcal{C}\ell_{\mathrm{Spin}_{1,3}^{e}}^{l}(M)$, which is a
vector bundle very different from $\mathcal{C}\ell(M,g)$. Their relation is
presented below, but before that we give the following definition.
\end{remark}

\begin{definition}
Let $I(M)$ be a subbundle of $\mathcal{C}\ell_{\mathrm{Spin}_{1,3}^{e}}%
^{l}(M)$ such that there exists a primitive idempotent $\mathbf{e}$ of
$\mathbb{R}_{1,3}$ (see, e.g., \cite{50}) with
\begin{equation}
R_{\mathbf{e}}\Psi=\Psi\mathbf{e}=\Psi
\end{equation}
for all $\Psi\in\sec I(M)\subset\sec\mathcal{C}\ell_{\mathrm{Spin}_{1,3}^{e}%
}^{l}(M).$ Then, $I(M)$ is called a subbundle of left ideal algebraic spinor
fields\textit{. Any} $\Psi\in\sec I(M)\subset\sec\mathcal{C}\ell
_{\mathrm{Spin}_{1,3}^{e}}^{l}(M)$ is called a left ideal algebraic spinor
field (\textit{LIASF)\label{LIASF}}. $I(M)$ can be thought of as a a real
spinor bundle for $M$ such that $\mathbf{M}$ in Eq.(\ref{1.7}) \ is a minimal
left ideal of $\mathbb{R}_{1,3}.$
\end{definition}

\begin{definition}
Two subbundles $I(M)$ and$\ I'(M)$ of
\textit{LIASF are said to be geometrically equivalent if the idempotents}
$e,e'\in\mathbb{R}_{1,3}$ (appearing in the previous definition) are related by an
element $u\in$ $\mathrm{Spin}_{1,3}^{e}$, i.e., $e'=ueu^{-1}$.
\end{definition}

\begin{definition}
\label{RsCbundle}The $\emph{right}$ real spin-Clifford bundle of $M$ is the
vector bundle
\begin{equation}
\mathcal{C}\ell_{\mathrm{Spin}_{1,3}^{e}}^{r}(M)=P_{\mathrm{Spin}_{1,3}^{e}%
}(M)\times_{r}\mathbb{R}_{1,3}. \label{RsCbundle'}%
\end{equation}
Sections of $\mathcal{C}\ell_{\mathrm{Spin}_{1,3}^{e}}^{r}(M)$ are called
right spin-Clifford fields
\end{definition}

In Eq.~(\ref{RsCbundle'}) $r$ refers to the representation of $\mathrm{Spin}%
_{1,3}^{e}$ on $\mathbb{R}_{1,3}$ given by $r(a)x=xa^{-1}$. As in the case for
the left real spin-Clifford bundle, there is a natural embedding
$P_{\mathrm{Spin}_{1,3}^{e}}(M)\hookrightarrow\mathcal{C}\ell_{\mathrm{Spin}%
_{1,3}^{e}}^{r}(M)$ which comes from the embedding $\mathrm{Spin}_{1,3}%
^{e}\hookrightarrow\mathbb{R}_{1,3}^{0}$. There exists also a natural left
action $L_{a}$ of $a\in\mathbb{R}_{1,3}$ on $\mathcal{C}\ell_{\mathrm{Spin}%
_{1,3}^{e}}^{r}(M)$. This will be proved in section 5.

\begin{definition}
Let $I^{\star}(M)$ be a subbundle of $\mathcal{C}\ell_{\mathrm{Spin}_{1,3}%
^{e}}^{r}(M)$ such that there exists a primitive idempotent element
$\mathbf{e}$ of $\mathbb{R}_{1,3}$ with%
\begin{equation}
L_{\mathbf{e}}\Psi=\mathbf{e}\Psi=\Psi\label{IM1}%
\end{equation}
for any $\Psi$\ $\in\sec I^{\star}(M)\subset\sec\mathcal{C}\ell_{\mathrm{Spin}%
_{1,3}^{e}}^{r}(M)$. Then, $I^{\star}(M)$ is called a subundle of right ideal
algebraic spinor fields. Any $\Psi\in\sec I^{\star}(M)\subset\sec
\mathcal{C}\ell_{\mathrm{Spin}_{1,3}^{e}}^{r}(M)$ is called a right ideal
algebraic spinor field (RIASF)\label{RIASF}. $I^{\star}(M)$ can be thought of
as a a real spinor bundle for $M$ such that $\mathbf{M}^{\star}$ in
Eq.~(\ref{1.7bis}) is a minimal right ideal of $\mathbb{R}_{1,3}.$\ \
\end{definition}

\begin{definition}
Two subbundles $I^{\star}(M)$ and $I^{\star}{} '(M)$ of
\textit{RIASF are said to be geometrically equivalent if the idempotents}
$e,e'\in\mathbb{R}_{1,3}$ (appearing in the previous definition) are related by an
element $u\in\mathrm{Spin}_{1,3}^{e}$, i.e., $e'=ueu^{-1}$.
\end{definition}

\begin{proposition}
\label{ciffxspincliff}In a spin manifold, we have
\[
\mathcal{C}\ell(M,g)=P_{\mathrm{Spin}_{1,3}^{e}}(M)\times_{\mathrm{Ad}%
}\mathbb{R}_{1,3}.
\]

\end{proposition}

\textbf{Proof. }Remember once again that the representation
\[
\mathrm{Ad}:\mathrm{Spin}_{1,3}^{e}\rightarrow\mathrm{Aut}(\mathbb{R}%
_{1,3})\quad\mathrm{Ad}_{u}a=uau^{-1}\qquad u\in\mathrm{Spin}_{1,3}^{e}%
\]
is such that $\mathrm{Ad}_{-1}=$ identity and so $\mathrm{Ad}$ descends to a
representation $\mathrm{Ad}^{\prime}$ of $\mathrm{SO}_{1,3}^{e}$ which we
considered above. It follows that when $P_{\mathrm{Spin}_{1,3}^{e}}(M)$ exists
$\mathcal{C}\ell(M,g)=P_{\mathrm{Spin}_{1,3}^{e}}(M)\times_{\mathrm{Ad}%
}\mathbb{R}_{1,3}$.$\blacksquare$

\subsection{Bundle of Modules over a Bundle of Algebras}

\begin{proposition}
\label{modovercliff}$S(M)$ (or $\mathcal{C}\ell_{\mathrm{Spin}_{1,3}^{e}}%
^{l}(M)$) is a bundle of (left) \emph{modules} over the bundle of algebras
$\mathcal{C}\ell(M,g)$. In particular, the sections of the spinor bundle
$S(M)$ (or $\mathcal{C}\ell_{\mathrm{Spin}_{1,3}^{e}}^{l}(M)$) constitute a
module over the sections of the Clifford bundle.
\end{proposition}

For the proof, see \cite{7}, page 97.

\begin{corollary}
\label{colorbom}Let $\Phi,\Psi\in\sec C\ell_{\mathrm{Spin}_{1,3}^{e}}^{l}(M)$
and $\Psi\neq0$. Then there exists $\mathit{\psi}\in\sec\mathcal{C}\ell(M,g)$
\ such that
\begin{equation}
\Psi=\mathit{\psi}\Phi. \label{3.20b}%
\end{equation}

\end{corollary}

\textbf{Proof. } It is an immediate consequence of
proposition~\ref{modovercliff}.$\blacksquare$

So, the corollary allows us to identify a \emph{correspondence} between some
sections of $\mathcal{C}\ell(M,g)$ and some sections of $I(M)$ or
$\mathcal{C}\ell_{\mathrm{Spin}_{1,3}^{e}}^{l}(M)$ once we fix a section on
$\mathcal{C}\ell_{\mathrm{Spin}_{1,3}^{e}}^{l}(M)$. This and other
correspondences will be essential for the theory of section 5.\ Once we
clarified the meaning of a bundle of modules $S(M)$ over a bundle of algebras
$\mathcal{C}\ell(M,g),$ we can give the following.

\begin{definition}
Two real left spinor bundles (see definition \ref{LsCbundle}) are equivalent
if and only if they are equivalent as bundles of $\mathcal{C}\ell(M,g)$
modules.
\end{definition}

\begin{remark}
Of course, geometrically equivalent\ real left spinor bundles are equivalent.
\end{remark}

\begin{remark}
\label{CSCBUNDLE}In what follows we denote the complexified left spin Clifford
bundle by $\mathbb{C}\ell_{\mathrm{Spin}_{1,3}^{e}}^{l}(M)=P_{\mathrm{Spin}%
_{1,3}^{e}}(M)\times_{l}\mathbb{C\otimes R}_{1,3}\equiv P_{\mathrm{Spin}%
_{1,3}^{e}}(M)\times_{r}\mathbb{R}_{4,1}$ and the complexified right spin
Clifford bundle by $\mathbb{C}\ell_{\mathrm{Spin}_{1,3}^{e}}^{r}%
(M)=P_{\mathrm{Spin}_{1,3}^{e}}(M)\times_{r}\mathbb{C\otimes R}_{1,3}\equiv
P_{\mathrm{Spin}_{1,3}^{e}}(M)\times_{r}\mathbb{R}_{4,1}$.
\end{remark}

\section{Dirac-Hestenes Spinor Fields}

Let $\mathbf{E}^{\mu}$, $\mu=0,1,2,3$ be the canonical basis of $\mathbb{R}%
^{1,3}\hookrightarrow\mathbb{R}_{1,3}$ which generates the algebra
$\mathbb{R}_{1,3}$. They satisfy the basic relation $\mathbf{E}^{\mu
}\mathbf{E}^{\nu}+\mathbf{E}^{\nu}\mathbf{E}^{\mu}=2\eta^{\mu\nu}$. As shown,
e.g., in \cite{50},%
\begin{equation}
\mathbf{e=}\frac{1}{2}(1+\mathbf{E}^{0})\in\mathbb{R}_{1,3}\label{dh1}%
\end{equation}
is a primitive idempotent of $\mathbb{R}_{1,3}$ and
\begin{equation}
\mathbf{f=}\frac{1}{2}(1+\mathbf{E}^{0})\frac{1}{2}(1+i\mathbf{E}%
^{2}\mathbf{E}^{1})\in\mathbb{C\otimes R}_{1,3}\label{dh2}%
\end{equation}
is a primitive idempotent of $\mathbb{C\otimes R}_{1,3}$. Now, let
$\mathbf{I=}\mathbb{R}_{1,3}\mathbf{e}$ and $\mathbf{I}_{\mathbb{C}%
}=\mathbb{C\otimes R}_{1,3}\mathbf{f}$ be, respectively, the minimal left
ideals of $\mathbb{R}_{1,3}$ and $\mathbb{C\otimes R}_{1,3}$ generated by
$\mathbf{e}$ and $\mathbf{f}$. Let $\mathbf{\phi=\phi e\in I}$ and
$\mathbf{\Psi=\Psi f\in I}_{\mathbb{C}}$. Then, any $\mathbf{\phi\in I}$ can
be written as
\begin{equation}
\mathbf{\phi=\psi e}\label{dh3}%
\end{equation}
with $\mathbf{\psi\in}\mathbb{R}_{1,3}^{0}$. Analogously, any $\mathbf{\Psi\in
I}_{\mathbb{C}}$ can be written as
\begin{equation}
\mathbf{\Psi=\psi e}\frac{1}{2}(1+i\mathbf{E}^{2}\mathbf{E}^{1}),\label{dh4}%
\end{equation}
with $\mathbf{\psi\in}\mathbb{R}_{1,3}^{0}$.

Now, $\mathbb{C\otimes R}_{1,3}\simeq\mathbb{R}_{4,1}$ $\simeq\mathbb{C(}4)$,
where $\mathbb{C(}4)$ is the algebra of the $4\times4$ complex matrices. We
can verify that
\begin{equation}
\left(
\begin{array}[c]{cccc}%
1 & 0 & 0 & 0\\
0 & 0 & 0 & 0\\
0 & 0 & 0 & 0\\
0 & 0 & 0 & 0
\end{array}
\right)  \label{dh5}%
\end{equation}
is a primitive idempotent of $\mathbb{C(}4)$ which is a matrix representation
of $\mathbf{f}$. In this way we can prove (as shown, e.g., in \cite{50}) that
there is a bijection between column spinors, i.e., elements of $\mathbb{C}%
^{4}$ (the complex $4$-dimensional vector space) and the elements of
$\mathbf{I}_{\mathbb{C}}$. All that, plus the definitions of the left real and
complex spin bundles and the subbundle $I(M)$ suggests the following.

\begin{definition}
Let $\Phi\in\sec I(M)\subset\sec\mathcal{C}\ell_{\mathrm{Spin}_{1,3}^{e}}%
^{l}(M)$ be as in definition \ref{LIASF}, i.e.,%
\begin{equation}
R_{\mathbf{e}}\Phi=\Phi\mathbf{e}=\Phi,\quad \mathbf{e}^{2}=\mathbf{e}=
\frac{1}{2}(1+\mathbf{E}^{0})\in\mathbb{R}_{1,3}.\label{dh5'}%
\end{equation}
A Dirac-Hestenes Spinor field (DHSF) associated with $\Phi$ is an \emph{even}
section\footnote{Note that it is meaningful to speak about even (or odd)
elements in $C\ell_{\mathrm{Spin}_{1,3}^{e}}^{l}(M)$ since $\mathrm{Spin}%
_{1,3}^{e}\subseteq\mathbb{R}_{1,3}^{0}.$
\par
{}} $\mathbf{\psi}$ of $\mathcal{C}\ell_{\mathrm{Spin}_{1,3}^{e}}^{l}(M)$ such
that
\begin{equation}
\Phi=\mathbf{\psi e}.\label{dh6}%
\end{equation}

\end{definition}

\begin{remark}
An equivalent definition of a DHSF is the following. Let $\Psi\in
\sec\mathbb{C}\ell_{\mathrm{Spin}_{1,3}^{e}}^{l}(M)$ be such that
\begin{equation}
R_{\mathbf{f}}\Psi=\Psi\mathbf{f}=\Psi,\quad \mathbf{f}^{2}=\mathbf{f}=
\frac{1}{2}(1+\mathbf{E}^{0})\frac{1}{2}(1+i\mathbf{E}^{2}\mathbf{E}^{1})
\in\mathbb{C}\otimes\mathbb{R}_{1,3}.
\end{equation}
Then, a DHSF associated with $\Psi$ is an even section $\mathbf{\psi}$ of
$\mathcal{C}\ell_{\mathrm{Spin}_{1,3}^{e}}^{l}(M)\subset\mathbb{C}%
\ell_{\mathrm{Spin}_{1,3}^{e}}^{l}(M)$ such that
\begin{equation}
\Psi=\mathbf{\psi f.}\label{dh7}%
\end{equation}

\end{remark}

\begin{remark}
In what follows, when we refer to a Dirac-Hestenes spinor field
\ $\mathbf{\psi}$ we omit for simplicity the wording associated with $\Phi$
(or $\Psi$). It is very important to observe that DHSF are not sums of even
multivector (tensor) fields although, under a local trivialization,
$\mathbf{\psi}$ $\in\sec C\ell_{\mathrm{Spin}_{1,3}^{e}}^{l}(M)$ is mapped on
an even element of\ $\mathbb{R}_{1,3}$. We emphasize that DHSF are particular
sections of a spinor bundle, not of the Clifford bundle. However, we show in
section 5 how these objects have representatives in the Clifford bundle.
\end{remark}

\section{The Many Faces of the Dirac Equation}

\subsection{ Dirac Equation for Covariant Dirac Fields}

As is well known \cite{12}, a \textit{covariant} Dirac spinor field is a
\ section $\mathbf{\Psi}\in\sec S_{c}(M)=P_{\mathrm{Spin}_{1,3}^{e}}%
(M)\times_{\mu_{l}}\mathbb{C}^{4}$. Let $(U=M,\Phi),\Phi(\mathbf{\Psi
})=(x,|\Psi(x)\rangle)$ be a global trivialization corresponding to a spin
frame $\Xi$\textit{\ }(definition~\ref{SPIN FRAME}), such that%
\begin{align}
s(\Xi) &  =\{e_{a}\}\in P_{\mathrm{SO}_{1,3}^{e}}(M),\text{ }e^{a}\in\sec
C\ell(M,g),\nonumber\\
e^{a}e^{b}+e^{b}e^{a} &  =2\eta^{ab},\quad a,b=0,1,2,3.\label{DE0}%
\end{align}
The usual Dirac equation in a Lorentzian spacetime for the spinor field
$\mathbf{\Psi}$ --- in interaction with an electromagnetic field\footnote{We
denote the space of sections of $p$-vectors by $\sec\bigwedge\nolimits^{p}%
(M)$.} $A\in\sec\bigwedge\nolimits^{1}(M)\subset\sec\mathcal{C}\ell(M,g)$ ---
is then \cite{15}%
\begin{equation}
i%
\mbox{\boldmath{$\gamma$}}%
^{a}(\mathbf{\nabla}_{e_{a}}^{s}+iqA_{a})|\Psi(x)\rangle-m|\Psi(x)\rangle
=0,\label{DE}%
\end{equation}
where \ $%
\mbox{\boldmath{$\gamma$}}%
^{a}\in\mathbb{C(}4)$, $a=0,1,2,3$, is a set of \textit{constant} Dirac
\textit{matrices} satisfying
\begin{equation}%
\mbox{\boldmath{$\gamma$}}%
^{a}%
\mbox{\boldmath{$\gamma$}}%
^{b}+%
\mbox{\boldmath{$\gamma$}}%
^{b}%
\mbox{\boldmath{$\gamma$}}%
^{a}=2\eta^{ab}.\label{DEC}%
\end{equation}

\subsection{ Dirac Equation in $\mathcal{C}\ell_{\mathrm{Spin}_{1,3}^{e}}%
^{l}(M,g)$}

Due to the one-to-one correspondence between \textit{ideal }sections of
$\mathbb{C}\ell_{\mathrm{Spin}_{1,3}^{e}}^{l}(M)$, $\mathcal{C}\ell
_{\mathrm{Spin}_{1,3}^{e}}^{l}(M)$ and of $S_{c}(M)$ as explained in section
3, we can \textit{translate} the Dirac equation (\ref{DE}) (for a covariant spinor
field) into an equation for a spinor field that is a section of $\mathbb{C}%
\ell_{\mathrm{Spin}_{1,3}^{e}}^{l}(M)$, and finally write an equivalent
equation for a \textit{DHSF }$\psi\in\sec C\ell_{\mathrm{Spin}_{1,3}^{e}}%
^{l}(M)$. In order to do that we introduce the spin-Dirac operator.

\begin{definition}
The (spin) Dirac operator acting on sections of $C\ell_{\mathrm{Spin}%
_{1,3}^{e}}^{l}(M)$ is the first order differential operator \cite{7}
\begin{equation}
D^{s}=e^{a}\mathbf{\nabla}_{e_{a}}^{s}\text{.}\label{DE02}%
\end{equation}
where $\{e^{a}\}$ is as in Eq.~(\ref{DE0}) and $\mathbf{\nabla}^{s}$ is the
spinor covariant derivative (see the Appendix).
\end{definition}

Now we give the details of the inverse translation. We start with the
following equation which we call the Dirac equation in $\mathcal{C}\ell
_{\mathrm{Spin}_{1,3}^{e}}^{l}(M)$, denoted \textit{DE}$\mathcal{C}\ell^{l}:$%
\begin{equation}
D^{s}\mathbf{\psi E}^{21}-m\mathbf{\psi E}^{0}-qA\mathbf{\psi}=0\label{DHE01}%
\end{equation}
where $\mathbf{\psi}\in\sec\mathcal{C}\ell_{\mathrm{Spin}_{1,3}^{e}}^{l}(M)$
is a \textit{DHSF }and $\mathbf{E}^{a}\in\mathbb{R}_{1,3}$ are such that
$\mathbf{E}^{a}\mathbf{E}^{b}+\mathbf{E}^{b}\mathbf{E}^{a}=2\eta^{ab}$.
Multiplying Eq.~(\ref{DHE01}) on the right by the idempotent $\mathbf{f=}%
\frac{1}{2}(1+\mathbf{E}^{0})\frac{1}{2}(1+i\mathbf{E}^{2}\mathbf{E}^{1}%
)\in\mathbb{C\otimes R}_{1,3}$, we get after some simple algebraic
manipulations the following equation for the (complex) left ideal
spin-Clifford field $\Psi=\mathbf{\psi f}\in\sec\mathbb{C}\ell_{\mathrm{Spin}%
_{1,3}^{e}}^{l}(M):$%
\begin{equation}
iD^{s}\Psi-m\Psi-qA\Psi=0.\label{DHE02}%
\end{equation}

Now we can easily show, using the methods of\ \cite{50}, that given any global
trivializations $(U=M,\Theta)$ and $(U=M,\Phi),$ of $\mathcal{C}\ell(M,g)$ and
$\mathcal{C}\ell_{\mathrm{Spin}_{1,3}^{e}}^{l}(M),$ there exists matrix
representations of the $\{e^{a}\}$ that are equal to the Dirac matrices
$\mathbf{\gamma}^{a}$ (appearing in Eq.~(\ref{DE})). In that way the
correspondence between Eqs.~(\ref{DE}), (\ref{DHE01}) and (\ref{DHE02}) is proved.

\begin{remark}
We emphasize at this point that we call Eq.~(\ref{DHE01}) the \textit{DE}%
$\mathcal{C}\ell^{l}$. It looks similar to the Dirac-Hestenes equation (on
Minkowski spacetime) discussed in \cite{50}, \ but it is indeed very different
from it, regarding its mathematical nature. The \textit{DE}$\mathcal{C}\ell^{l}$ is
an intrinsic equation satisfied by a legitimate spinor field, namely a
\textit{DHSF }$\mathbf{\psi}\in \sec\mathcal{C}\ell_{\mathrm{Spin}_{1,3}^{e}}^{l}(M)$.
The question naturally arises: May we write an equation with the same mathematical
information of Eq.~(\ref{DHE01}) but satisfied by objects living on the Clifford
bundle $\mathcal{C}\ell(M,g)$ of an arbitrary Lorentzian spacetime, admitting a
spin structure? In the next section we show that the answer to that question is yes.
\end{remark}

\subsection{Electromagnetic Gauge Invariance of the \textit{DE}$\mathcal{C}%
\ell^{l}$}

\begin{proposition}
The \textit{DE}$\mathcal{C}\ell^{l}$ is invariant under electromagnetic gauge
transformations%
\begin{align}
\mathbf{\psi}  &  \mapsto\mathbf{\psi}^{\prime}=\mathbf{\psi}e^{q\mathbf{E}%
^{21}\chi},\label{GI1}\\
A  &  \mapsto A+\partial\chi,\label{GI2}\\
\omega_{e_{a}}  &  \mapsto\omega_{e_{a}}\label{GI3}\\
\mathbf{\psi,\psi}^{\prime}  &  \in\sec\mathcal{C}\ell_{\mathrm{Spin}%
_{1,3}^{e}}^{l}(M)\\
A  &  \in\sec\bigwedge\nolimits^{1}(M)\subset\sec\mathcal{C}\ell(M,g)
\end{align}
with $\mathbf{\psi,\psi}^{\prime}$ DHSF, and where $\chi
:M\rightarrow\mathbb{R\subset R}_{1,3}$ is a gauge function.
\end{proposition}

\textbf{Proof.} The proof is obtained by direct verification.$\blacksquare$

\begin{remark}
We note that, for the \textit{DE}$\mathcal{C}\ell^{l}$, local rotations and
electromagnetic gauge transformations are very different mathematical
transformations, without any obvious geometrical link between them,
differently of what seems to be the case for the Dirac-Hestenes equation,
which is studied in the next section.
\end{remark}

\section{The Dirac-Hestenes Equation (\textit{DHE})}

We obtained above a Dirac equation, which we called \textit{DE}$\mathcal{C}%
\ell^{l}$, describing the motion of spinor fields represented by sections
$\mathbf{\Psi}$ of $C\ell_{\mathrm{Spin}_{1,3}^{e}}^{l}(M,g)$ in interaction
with an electromagnetic field $A\in\sec C\ell(M,g),$
\begin{equation}
D^{s}\mathbf{\Psi E}^{21}-qA\mathbf{\Psi}=m\mathbf{\Psi E}^{0}%
,\label{DE for Psi in Cl_left}%
\end{equation}
where $D^{s}=e^{a}\mathbf{\nabla}_{e_{a}}^{s}$, $\{e^{a}\}$ is given by
Eq.~(\ref{DE0}), $\mathbf{\nabla}_{e_{a}}^{s}$ is the natural spinor covariant
derivative acting on $\sec C\ell_{\mathrm{Spin}_{1,3}^{e}}^{l}(M,g)$ (see the
Appendix), and $\{\mathbf{E}^{a}\}\in\mathbb{R}^{1,3}\subseteq\mathbb{R}%
_{1,3}$ is such that $\mathbf{E}^{a}\mathbf{E}^{b}+\mathbf{E}^{b}%
\mathbf{E}^{a}=2\eta^{ab}$. As we already mentioned, although Eq.
(\ref{DE for Psi in Cl_left}) is written in a kind of Clifford bundle (i.e.
$C\ell_{\mathrm{Spin}_{1,3}^{e}}^{l}(M,g)$), it does not suffer from the
inconsistency of representing spinors as pure differential forms and, in fact,
the object $\mathbf{\Psi}$ behaves as it should under Lorentz transformations.

As a matter of fact, Eq.~(\ref{DE for Psi in Cl_left}) can be thought of as a
mere \textit{rewriting} of the usual Dirac equation, where the role of the
constant gamma matrices is undertaken by the constant elements $\{\mathbf{E}%
^{a}\}$ in $\mathbb{R}_{1,3}$ and by the set $\{e^{a}\}$. In this way,
Eq.~(\ref{DE for Psi in Cl_left}) is \emph{not} a kind of Dirac-Hestenes
equation as discussed, e.g., in \cite{50}. It suffices to say that (i) the
state of the electron, represented by $\mathbf{\Psi}$, is not a
\textit{Clifford field} and (ii) the $\mathbf{E}^{a}$'s are just
\textit{constant} elements of $\mathbb{R}_{1,3}$ and not sections of vectors
in $\mathcal{C}\ell(M,g)$. Nevertheless, as we show in the following, Eq.
(\ref{DE for Psi in Cl_left}) does lead to a multivector Dirac equation once
we carefully employ the theory of right and left actions on the various
Clifford bundles introduced earlier. It is the multivector
equation\footnote{Of course, we can write an equivalent multiform equation.}
to be derived below that we call the \textit{DHE}. We shall need several
preliminary results that we collect in the next two subsections.

\subsection{The Various Natural Actions on the Vector Bundles Associated to
$P_{\mathrm{Spin}_{1,3}^{e}}(M)$}

Recall that, when $M$ is a spin manifold the following occurs.

(i) The elements of $\mathcal{C}\ell(M,g)=P_{\mathrm{Spin}_{1,3}^{e}}%
(M)\times_{Ad}\mathbb{R}_{1,3}$ are equivalence classes $\left[  (p,a)\right]
$ of pairs $(p,a),$ where $p\in P_{\mathrm{Spin}_{1,3}^{e}}(M)$,
$a\in\mathbb{R}_{1,3}$ and $(p,a)\sim(p^{\prime},a^{\prime})$ $\Leftrightarrow
p^{\prime}=pu^{-1},$ $a^{\prime}=uau^{-1}$, for some $u\in\mathrm{Spin}%
_{1,3}^{e}$.

(ii) The elements of $\mathcal{C}\ell_{\mathrm{Spin}_{1,3}^{e}}^{l}(M)$ are
equivalence classes of pairs $(p,a),$ where $p\in P_{\mathrm{Spin}_{1,3}^{e}%
}(M)$, $a\in\mathbb{R}_{1,3}$ and $(p,a)\sim(p^{\prime},a^{\prime})$
$\Leftrightarrow p^{\prime}=pu^{-1},$ $a^{\prime}=ua$, for some $u\in
\mathrm{Spin}_{1,3}^{e}$.

(iii) The elements of $\mathcal{C}\ell_{\mathrm{Spin}_{1,3}^{e}}^{r}(M)$ are
equivalence classes of pairs $(p,a),$ where $p\in P_{\mathrm{Spin}_{1,3}^{e}%
}(M)$, $a\in\mathbb{R}_{1,3}$ and $(p,a)\sim(p^{\prime},a^{\prime})$
$\Leftrightarrow p^{\prime}=pu^{-1},$ $a^{\prime}=au^{-1}$, for some
$u\in\mathrm{Spin}_{1,3}^{e}$.

In this way, it is possible to define the following natural actions on these
associated bundles.

\begin{proposition}
There is a natural right action of $\mathbb{R}_{1,3}$ on $\mathcal{C}%
\ell_{\mathrm{Spin}_{1,3}^{e}}^{l}(M)$ and a natural left action of
$\mathbb{R}_{1,3}$ on $\mathcal{C}\ell_{\mathrm{Spin}_{1,3}^{e}}^{r}%
(M,g)$.\label{R_13 on Cl_left}
\end{proposition}

\textbf{Proof.} Given $b\in\mathbb{R}_{1,3}$ and $\alpha\in\mathcal{C}%
\ell_{\mathrm{Spin}_{1,3}^{e}}^{l}(M,g),$ select a representative $(p,a)$ for
$\alpha$ and define $\alpha b:=\left[  (p,ab)\right]  \in \mathcal{C}%
\ell_{\mathrm{Spin}_{1,3}^{e}}^{l}(M,g).$ If another
representative $(pu^{-1},ua)$ is chosen for $\alpha,$ we have $(pu^{-1}%
,uab)\sim(p,ab)$ and thus $\alpha b$ is a well-defined element of
$\mathcal{C}\ell_{\mathrm{Spin}_{1,3}^{e}}^{l}(M)$.$\blacksquare$

Let us denote the space of $\mathbb{R}_{1,3}$-valued smooth functions on $M$
by $\mathcal{F}(M,\mathbb{R}_{1,3})$. Then, the above proposition immediately
yields the following.

\begin{corollary}
There is a natural right action of $\mathcal{F}(M,\mathbb{R}_{1,3})$ on
$\sec\mathcal{C}\ell_{\mathrm{Spin}_{1,3}^{e}}^{l}(M)$ and a natural left
action of $\mathcal{F}(M,\mathbb{R}_{1,3})$ on $\sec\mathcal{C}\ell
_{\mathrm{Spin}_{1,3}^{e}}^{r}(M,g)$.\label{C(R_13) on sec(Cl_left)}
\end{corollary}

\begin{proposition}
There is a natural left action of $\sec\mathcal{C}\ell(M,g)$ on $\sec
\mathcal{C}\ell_{\mathrm{Spin}_{1,3}^{e}}^{l}(M)$ and a natural right action
of $\sec\mathcal{C}\ell(M,g)$ on $\sec\mathcal{C}\ell_{\mathrm{Spin}_{1,3}%
^{e}}^{r}(M)$.\label{sec(Cl) on sec(Cl_left)}
\end{proposition}

\textbf{Proof.} Given $\alpha\in\sec\mathcal{C}\ell(M,g)$ and $\beta\in
\sec\mathcal{C}\ell_{\mathrm{Spin}_{1,3}^{e}}^{l}(M,g),$ select
representatives $(p,a)$ for $\alpha(x)$ and $(p,b)$ for $\beta(x)$ (with
$p\in\pi^{-1}(x)$) and define $(\alpha\beta)(x):=\left[  (p,ab)\right]
\in\mathcal{C}\ell_{\mathrm{Spin}_{1,3}^{e}}^{l}(M,g).$ If alternative
representatives $(pu^{-1},uau^{-1})$ and $(pu^{-1},ub)$ are chosen for
$\alpha(x)$ and $\beta(x),$ we have%
\[
(pu^{-1},uau^{-1}ub)=(pu^{-1},uab)\sim(p,ab)
\]
and thus $(\alpha\beta)(x)$ is a well-defined element of $\mathcal{C}%
\ell_{\mathrm{Spin}_{1,3}^{e}}^{l}(M,g)$.$\blacksquare$

\begin{proposition}
\label{sec(Cl_left) x sec(Cl_right)}There is a natural pairing%
\[
\sec\mathcal{C}\ell_{\mathrm{Spin}_{1,3}^{e}}^{l}(M)\times\sec\mathcal{C}%
\ell_{\mathrm{Spin}_{1,3}^{e}}^{r}(M)\rightarrow\sec\mathcal{C}\ell(M,g).
\]

\end{proposition}

\textbf{Proof.} Given $\alpha\in\sec\mathcal{C}\ell_{\mathrm{Spin}_{1,3}^{e}%
}^{l}(M)$ and $\beta\in\sec\mathcal{C}\ell_{\mathrm{Spin}_{1,3}^{e}}^{r}(M),$
select representatives $(p,a)$ for $\alpha(x)$ and $(p,b)$ for $\beta(x)$
(with $p\in\pi^{-1}(x)$) and define $(\alpha\beta)(x):=\left[  (p,ab)\right]
\in\mathcal{C}\ell(M,g)$. If alternative representatives $(pu^{-1},ua)$ and
$(pu^{-1},bu^{-1})$ are chosen for $\alpha(x)$ and $\beta(x),$ we have
$(pu^{-1},uabu^{-1})\sim(p,ab)$ and thus $(\alpha\beta)(x)$ is a well-defined
element of $C\ell(M,g)$.$\blacksquare$

\begin{proposition}
\label{sec(CL_right) on sec(Cl_left)}There is a natural pairing%
\[
\sec\mathcal{C}\ell_{\mathrm{Spin}_{1,3}^{e}}^{r}(M)\times\sec\mathcal{C}%
\ell_{\mathrm{Spin}_{1,3}^{e}}^{l}(M)\rightarrow\mathcal{F}(M,\mathbb{R}%
_{1,3}).
\]

\end{proposition}

\textbf{Proof.} Given $\alpha\in\sec\mathcal{C}\ell_{\mathrm{Spin}_{1,3}^{e}%
}^{r}(M)$ and $\beta\in\sec\mathcal{C}\ell_{\mathrm{Spin}_{1,3}^{e}}^{l}(M),$
select representatives $(p,a)$ for $\alpha(x)$ and $(p,b)$ for $\beta(x)$
(with $p\in\pi^{-1}(x)$) and define $(\alpha\beta)(x):=ab\in\mathbb{R}_{1,3}$.
If alternative representatives $(pu^{-1},au^{-1})$ and $(pu^{-1},ub)$ are
chosen for $\alpha(x)$ and $\beta(x),$ we have $au^{-1}ub=ab$ and thus
$(\alpha\beta)(x)$ is a well-defined element of $\mathbb{R}_{1,3}%
$.$\blacksquare$

\subsection{Fiducial Sections Associated with a Spin Frame}

\label{sec fid}

We start by exploring the possibility of defining \textquotedblleft unit
sections\textquotedblright\ on the various vector bundles associated with the
principal bundle $P_{\mathrm{Spin}_{1,3}^{e}}(M)$. It immediately follows from
the definition given by Eq.~(\ref{1.1}) that the unit section $\mathbf{1}%
\in\sec\mathcal{C}\ell(M,g)$, given by $x\mapsto1\in\mathcal{C}\ell
(T_{x}M,g_{x})$, is certainly well defined. For future reference, let us
consider how this can also be seen from the associated bundle structure of
$P_{\mathrm{Spin}_{1,3}^{e}}(M)\times_{ad}\mathbb{R}_{1,3}$.

Let%
\begin{equation}
\Phi_{i}:\mathbf{\pi}^{-1}(U_{i})\rightarrow U_{i}\times\mathrm{Spin}%
_{1,3}^{e},\quad\Phi_{j}:\mathbf{\pi}^{-1}(U_{j})\rightarrow U_{j}%
\times\mathrm{Spin}_{1,3}^{e}\nonumber
\end{equation}
be two local trivializations for $P_{\mathrm{Spin}_{1,3}^{e}}(M),$ with%
\[
\Phi_{i}(u)=(\pi(u)=x,\phi_{i,x}(u)),\quad\Phi_{j}(u)=(\pi(u)=x,\phi
_{j,x}(u)).
\]
Recall that the corresponding transition function $g_{ij}:U_{i}\cap
U_{j}\rightarrow\mathrm{Spin}_{1,3}^{e}$ is then given by%
\[
g_{ij}(x)=\phi_{i,x}(u)\circ\phi_{j,x}(u)^{-1},
\]
which does not depend on $u$.

\begin{proposition}
$\mathcal{C}\ell(M,g)$ has a naturally defined global unit section.
\end{proposition}

\textbf{Proof. \ }For the associated bundle $\mathcal{C}\ell
(M,g)=P_{\mathrm{Spin}_{1,3}^{e}}(M)\times_{\mathrm{Ad}}\mathbb{R}_{1,3}$, the
transition functions corresponding to local trivializations
\begin{equation}
\Psi_{i}:\mathbf{\pi}_{c}^{-1}(U_{i})\rightarrow U_{i}\times\mathbb{R}%
_{1,3}\text{,\quad}\Psi_{j}:\mathbf{\pi}_{c}^{-1}(U_{j})\rightarrow
U_{j}\times\mathbb{R}_{1,3},
\end{equation}
are given by $h_{ij}(x)=Ad_{g_{ij}(x)}$. Define the local sections
\begin{equation}
\mathbf{1}_{i}(x)=\Psi_{i}^{-1}(x,1),\quad\mathbf{1}_{j}(x)=\Psi_{j}%
^{-1}(x,1),\label{?1}%
\end{equation}
where $1$ is the unit element of $\mathbb{R}_{1,3}$. Since $h_{ij}%
(x)\cdot1=Ad_{g_{ij}(x)}(1)=g_{ij}(x)1g_{ij}(x)^{-1}=1$, we see that the
expressions above uniquely define a global section $\mathbf{1\in}%
\mathcal{C}\ell(M,g)$ with $\mathbf{1}|_{U_{i}}=\mathbf{1}_{i}$.$\blacksquare$

It is clear that such a result can be immediately generalized for the Clifford
bundle $\mathcal{C}\ell_{p,q}(M,g)$, of any $n$-dimensional manifold endowed
with a metric of arbitrary signature $(p,q)$ (where $n=p+q$). Now, we observe
also that the left (and also the right) spin-Clifford bundle can be
generalized in an obvious way for any spin manifold of arbitrary finite
dimension $n=p+q$, with a metric of arbitrary signature $(p,q)$. However,
another important difference between $\mathcal{C}\ell(M,g)$ and $\mathcal{C}%
\ell_{\mathrm{Spin}_{p,q}^{e}}^{l}(M)$ or $\mathcal{C}\ell_{\mathrm{Spin}%
_{1,3}^{e}}^{r}(M,g)$ is that these latter bundles only admit a global unit
section if they are \textit{trivial}.

\begin{proposition}
\label{IDENT?}There exists a unit section on $\mathcal{C}\ell_{\mathrm{Spin}%
_{p,q}^{e}}^{r}(M)$ (and also on $\mathcal{C}\ell_{\mathrm{Spin}_{p,q}^{e}%
}^{l}(M)$) if and only if $P_{\mathrm{Spin}_{p,q}^{e}}(M)$ is trivial.
\end{proposition}

\textbf{Proof.} We show the necessity for the case of $\mathcal{C}%
\ell_{\mathrm{Spin}_{p,q}^{e}}^{r}(M)$,\footnote{The proof for the case of
$\mathcal{C}\ell_{\mathrm{Spin}_{p,q}^{e}}^{l}(M)$ is analogous.} the
sufficiency is trivial. \ For $\mathcal{C}\ell_{\mathrm{Spin}_{p,q}^{e}}%
^{r}(M)$, the transition functions corresponding to local trivializations
\begin{equation}
\Omega_{i}:\mathbf{\pi}_{sc}^{-1}(U_{i})\rightarrow U_{i}\times\mathbb{R}%
_{p,q}\text{,\quad}\Omega_{j}:\mathbf{\pi}_{sc}^{-1}(U_{j})\rightarrow
U_{j}\times\mathbb{R}_{p,q},
\end{equation}
are given by $k_{ij}(x)=R_{g_{ij}(x)}$, with $R_{a}:$ $\mathbb{R}%
_{p,q}\rightarrow\mathbb{R}_{p,q},x\mapsto xa^{-1}$. Let $1$ be the unit
element of $\mathbb{R}_{1,3}$. A unit section in $\mathcal{C}\ell
_{\mathrm{Spin}_{p,q}^{e}}^{r}(M)$ --- if it exists --- is written in terms of
these two local trivializations as
\begin{equation}
\mathbf{1}_{i}^{r}(x)=\Omega_{i}^{-1}(x,1),\quad\mathbf{1}_{j}^{r}%
(x)=\Omega_{j}^{-1}(x,1),
\end{equation}
and we must have $\mathbf{1}_{i}^{r}(x)=\mathbf{1}_{j}^{r}(x)$ $\forall x\in
U_{i}\cap U_{j}$. As $\Omega_{i}(\mathbf{1}_{i}^{r}(x))=(x,1)=$ $\Omega
_{j}(\mathbf{1}_{j}^{r}(x))$, we have $\mathbf{1}_{i}^{r}(x)=\mathbf{1}%
_{j}^{r}(x)$ $\Leftrightarrow1=k_{ij}(x)\cdot1\Leftrightarrow1=1g_{ij}%
(x)^{-1}\Leftrightarrow g_{ij}(x)=1$. This proves the
proposition.$\blacksquare$

\begin{remark}
For general spin manifolds, the bundle $P_{\mathrm{Spin}_{p,q}^{e}}(M)$ is not
necessarily trivial for arbitrary $(p,q)$, but Geroch's theorem (remark
\ref{geroch rem 2}) warrants that, for the special case $(p,q)=(1,3)$ with $M$
noncompact, $P_{\mathrm{Spin}_{1,3}^{e}}(M)$ is trivial. By the above
proposition, we then see that $\mathcal{C}\ell_{\mathrm{Spin}_{1,3}^{e}}%
^{r}(M)$ and also $\mathcal{C}\ell_{\mathrm{Spin}_{1,3}^{e}}^{l}(M)$
have\ global \textquotedblleft unit sections\textquotedblright. It is most
important to note, however, that each different choice of a (global)
trivialization $\Omega_{i}$ on $\mathcal{C}\ell_{\mathrm{Spin}_{1,3}^{e}}%
^{r}(M)$ (respectively, $\mathcal{C}\ell_{\mathrm{Spin}_{p,q}^{e}}^{l}(M)$)
induces a different global unit section $\mathbf{1}_{i}^{r}$ (respectively,
$\mathbf{1}_{i}^{l}$). Therefore, even in this case there is no canonical unit
section on $\mathcal{C}\ell_{\mathrm{Spin}_{1,3}^{e}}^{r}(M,g)$ (respectively,
on $\mathcal{C}\ell_{\mathrm{Spin}_{1,3}^{e}}^{l}(M,g)$).
\end{remark}

By remark \ref{geroch rem 2}, when the (noncompact) spacetime $M$ is a spin
manifold, the bundle $P_{\mathrm{Spin}_{1,3}^{e}}(M)$ admits global sections.
With this in mind, let us fix a spin frame $\Xi$ for $M$. This induces a
global trivialization for $P_{\mathrm{Spin}_{1,3}^{e}}(M),$ which we denote by
$\Phi_{\Xi}:P_{\mathrm{Spin}_{1,3}^{e}}(M)\rightarrow M\times\mathrm{Spin}%
_{1,3}^{e},$ with $\Phi_{\Xi}^{-1}(x,1)=\Xi(x)$. As we show in the following,
the spin frame $\Xi$ can also be used to induce certain fiducial global
sections on the various vector bundles associated with $P_{\mathrm{Spin}%
_{1,3}^{e}}(M)$:

\subparagraph{(i) $\mathcal{C}\ell(M,g)$}

Let $\{\mathbf{E}^{a}\}$ be a fixed orthonormal basis of $\mathbb{R}%
^{1,3}\subseteq\mathbb{R}_{1,3}$ (which can be thought of as the
canonical basis of $\mathbb{R}^{1,3}$). We define basis sections in
$\mathcal{C}\ell(M,g)=P_{\mathrm{Spin}_{1,3}^{e}}(M)\times_{Ad}\mathbb{R}%
_{1,3}$ by $e_{a}(x)=\left[  (\Xi(x),\mathbf{E}_{a})\right]  $. Of course,
this induces a multivector basis $\{e_{I}(x)\}$ for each $x\in M$. Note that a
more precise notation for $e_{a}$ would be, for instance, $e_{a}^{(\Xi)}$.

\subparagraph{(ii) $\mathcal{C}\ell_{\mathrm{Spin}_{1,3}^{e}}^{l}(M)$}

Let $\mathbf{1}_{\Xi}^{l}\in\sec\mathcal{C}\ell_{\mathrm{Spin}_{1,3}^{e}}%
^{l}(M)$ be defined by $\mathbf{1}_{\Xi}^{l}(x)=\left[  (\Xi(x),1)\right]  $.
Then the natural right action of $\mathbb{R}_{1,3}$ on $\mathcal{C}%
\ell_{\mathrm{Spin}_{1,3}^{e}}^{l}(M)$ leads to $\mathbf{1}_{\Xi}%
^{l}(x)a=\left[  (\Xi(x),a)\right]  $ for all $a\in\mathbb{R}_{1,3}$. It
follows from corollary \ref{C(R_13) on sec(Cl_left)} that an arbitrary section
$\alpha\in\sec\mathcal{C}\ell_{\mathrm{Spin}_{1,3}^{e}}^{l}(M)$ can be written
as $\alpha=\mathbf{1}_{\Xi}^{l}f$, with $f\in\mathcal{F}(M,\mathbb{R}_{1,3})$.

\subparagraph{(iii) $\mathcal{C}\ell_{\mathrm{Spin}_{1,3}^{e}}^{r}(M,g)$}

Let $\mathbf{1}_{\Xi}^{r}\in\sec\mathcal{C}\ell_{\mathrm{Spin}_{1,3}^{e}}%
^{r}(M,g)$ be defined by $\mathbf{1}_{\Xi}^{r}(x)=\left[  (\Xi(x),1)\right]
$. Then the natural left action of $\mathbb{R}_{1,3}$ on $\mathcal{C}%
\ell_{\mathrm{Spin}_{1,3}^{e}}^{r}(M)$ leads to $a\mathbf{1}_{\Xi}%
^{r}(x)=\left[  (\Xi(x),a)\right]  $ for all $a\in\mathbb{R}_{1,3}$. It
follows from corollary \ref{C(R_13) on sec(Cl_left)} that an arbitrary section
$\alpha\in\sec\mathcal{C}\ell_{\mathrm{Spin}_{1,3}^{e}}^{r}(M)$ can be written
as $\alpha=f\mathbf{1}_{\Xi}^{r}$, with $f\in\mathcal{F}(M,\mathbb{R}_{1,3}%
)$.\bigskip

Now recall (definition \ref{spin structure}) that a spin structure on $M$ is a
2-1 bundle map $s:P_{\mathrm{Spin}_{1,3}^{e}}(M)\rightarrow P_{\mathrm{SO}%
_{1,3}^{e}}(M)$ such that $s(pu)=s(p)Ad_{u},\ \forall p\in P_{\mathrm{Spin}%
_{1,3}^{e}}(M),$ $u\in\mathrm{Spin}_{1,3}^{e}$, where $Ad:\mathrm{Spin}%
_{1,3}^{e}\rightarrow\mathrm{SO}_{1,3}^{e},$ $Ad_{u}:x\mapsto uxu^{-1}$. We
see that the specification of the global section in the case (i) above is
compatible with the Lorentz frame $\{e_{a}\}=s(\Xi)$ assigned by $s$. More
precisely, for each $x\in M$, the element $s(\Xi(x))\in P_{\mathrm{SO}%
_{1,3}^{e}}(M)$ is to be regarded as a proper isometry $s(\Xi(x)):\mathbb{R}%
^{1,3}\rightarrow T_{x}M$, so that $e_{a}(x):=s(p)\cdot\mathbf{E}_{a}$ yields
a Lorentz frame $\{e_{a}\}$ on $M$, which we denoted by $s(\Xi)$. On the other
hand, $\mathcal{C}\ell(M,g)$ is isomorphic to $P_{\mathrm{Spin}_{1,3}^{e}%
}(M)\times_{Ad}\mathbb{R}_{1,3}$, and we can always arrange things so that
$e_{a}(x)$ is represented in this bundle as $e_{a}(x)=\left[  (\Xi
(x),\mathbf{E}_{a})\right]  .$ In fact, all we have to do is to verify that
this identification is covariant under a change of frames. To see that, let
$\Xi^{\prime}\in\sec P_{\mathrm{Spin}_{1,3}^{e}}(M)$ be another spin frame on
$M$. From the principal bundle structure of $P_{\mathrm{Spin}_{1,3}^{e}}(M)$,
we know that, for each $x\in M$, there exists (a unique) $u(x)\in
\mathrm{Spin}_{1,3}^{e}$ such that $\Xi^{\prime}(x)=\Xi(x)u(x)$. If we define,
as above, $e_{a}^{\prime}(x)=s(\Xi^{\prime}(x))\cdot\mathbf{E}_{a}$, then
$e_{a}^{\prime}(x)=s(\Xi(x)u(x))\cdot\mathbf{E}_{a}=s(\Xi(x))Ad_{u(x)}%
\cdot\mathbf{E}_{a}=\left[  (\Xi(x),Ad_{u(x)}\cdot\mathbf{E}_{a})\right]
=\left[  (\Xi(x)u(x),\mathbf{E}_{a})\right]  =\left[  (\Xi^{\prime
}(x),\mathbf{E}_{a})\right]  $, which proves our claim.

\begin{proposition}
\label{E_a in terms of e_a}%
\begin{align*}
\text{(i) }\mathbf{E}_{a}  &  =\mathbf{1}_{\Xi}^{r}(x)e_{a}(x)\mathbf{1}_{\Xi
}^{l}(x)\text{, }\forall x\in M,\\
\text{(ii) }\mathbf{1}_{\Xi}^{l}\mathbf{1}_{\Xi}^{r}  &  =1\in\mathcal{C}%
\ell(M,g),\\
\text{(iii) }\mathbf{1}_{\Xi}^{r}\mathbf{1}_{\Xi}^{l}  &  =1\in\mathbb{R}%
_{1,3}.
\end{align*}

\end{proposition}

\textbf{Proof.} This follows from the form of the various actions defined in
propositions \ref{R_13 on Cl_left}-\ref{sec(CL_right) on sec(Cl_left)}. For
example, for each $x\in M,$ we have $\mathbf{1}_{\Xi}^{r}(x)e_{a}%
(x)=[(\Xi(x),1\mathbf{E}_{a})]=[(\Xi(x),\mathbf{E}_{a})]\in\sec\mathcal{C}%
\ell_{\mathrm{Spin}_{1,3}^{e}}^{r}(M)$ (from proposition
\ref{sec(Cl) on sec(Cl_left)}). Then, it follows from proposition
\ref{sec(CL_right) on sec(Cl_left)} that $\mathbf{1}_{\Xi}^{r}(x)e_{a}%
(x)\mathbf{1}_{\Xi}^{l}(x)=\mathbf{E}_{a}1=\mathbf{E}_{a}$ $\forall x\in
M$.$\blacksquare$

Let us now consider how the various global sections defined above transform
when the spin frame $\Xi$ is changed. Let $\Xi^{\prime}\in\sec
P_{\mathrm{Spin}_{1,3}^{e}}(M)$ be another spin frame with $\Xi^{\prime
}(x)=\Xi(x)u(x)$, where $u(x)\in\mathrm{Spin}_{1,3}^{e}$. Let $e_{a}$,
$\mathbf{1}_{\Xi}^{r}$, $\mathbf{1}_{\Xi}^{l}$ and $e_{a}^{\prime}$,
$\mathbf{1}_{\Xi^{\prime}}^{r}$, $\mathbf{1}_{\Xi^{\prime}}^{l}$ be the global
sections, respectively, defined by $\Xi$ and $\Xi^{\prime}$ (as above). We
then have the following.

\begin{proposition}
\label{transformation of 1_right}Let $\Xi,\Xi^{\prime}$ be two spin frames
related by $\Xi^{\prime}=\Xi u$, where $u:M\rightarrow\mathrm{Spin}_{1,3}^{e}%
$. Then%
\begin{align}
(i)\text{ }e_{a}^{\prime} &  =Ue_{a}U^{-1}\nonumber\\
(ii)\text{ }\mathbf{1}_{\Xi^{\prime}}^{l} &  =\mathbf{1}_{\Xi}^{l}%
u=U\mathbf{1}_{\Xi}^{l},\nonumber\\
(iii)\text{ }\mathbf{1}_{\Xi^{\prime}}^{r} &  =u^{-1}\mathbf{1}_{\Xi}%
^{r}=\mathbf{1}_{\Xi}^{r}U^{-1},\label{transforms}%
\end{align}
where $U\in\sec\mathcal{C}\ell(M,g)$ is the Clifford field associated with $u$
by $U(x)=[(\Xi(x),u(x))]$. Also, in (ii) and (iii), $u$ and $u^{-1}$,
respectively, act on $\mathbf{1}_{\Xi}^{l}\in\sec\mathcal{C}\ell
_{\mathrm{Spin}_{1,3}^{e}}^{l}(M)$ and $\mathbf{1}_{\Xi}^{r}\in\sec
\mathcal{C}\ell_{\mathrm{Spin}_{1,3}^{e}}^{r}(M)$ according to proposition
\ref{C(R_13) on sec(Cl_left)}.
\end{proposition}

\textbf{Proof.} (i) We have%
\begin{align}
e_{a}^{\prime}(x)  &  =[(\Xi^{\prime}(x),\mathbf{E}_{a})]=[(\Xi
(x)u(x),\mathbf{E}_{a})]\nonumber\\
&  =[(\Xi(x),u(x)\mathbf{E}_{a}u(x)^{-1})]\nonumber\\
&  =[(\Xi(x),u(x))][(\Xi(x),\mathbf{E}_{a})][(\Xi(x),u(x)^{-1})]\nonumber\\
&  =U(x)e_{a}(x)U(x)^{-1}.
\end{align}

\noindent

(iii) It follows from proposition \ref{sec(Cl) on sec(Cl_left)} that%
\begin{align}
\mathbf{1}_{\Xi^{\prime}}^{r}(x) &  =\left[  (\Xi^{\prime}(x),1)\right]
=\left[  (\Xi(x)u(x),1)\right]  \nonumber\\
&  =\left[  (\Xi(x),1u(x)^{-1})\right]  =\left[  (\Xi(x),u(x)^{-1})\right]
=u(x)^{-1}\mathbf{1}_{\Xi}^{r}(x),
\end{align}
where in the last step we used proposition \ref{C(R_13) on sec(Cl_left)} and
the fact that $\mathbf{1}_{\Xi}^{r}(x)=\left[  (\Xi(x),1)\right]  $. To
demonstrate the second part, note that%
\begin{align}
u^{-1}(x)\mathbf{1}_{\Xi}^{r}(x) &  =\left[  (\Xi(x),u(x)^{-1})\right]
\nonumber\\
&  =\left[  (\Xi(x),1u(x)^{-1})\right]  =\left[  (\Xi(x),1)\right]  \left[
(\Xi(x),u(x)^{-1})\right]  \nonumber\\
&  =\mathbf{1}_{\Xi}^{r}(x)U^{-1}(x),
\end{align}
for all $x\in M.$ It is important to note that in the last step we have a
product between an element of $\mathcal{C}\ell_{\mathrm{Spin}_{1,3}^{e}}%
^{r}(M)$ (i.e., $\left[  (\Xi(x),1)\right]  $) and an element of
$\mathcal{C}\ell(M,g)$ (i.e., $\left[  (\Xi(x),u(x)^{-1})\right]
$).$\blacksquare$

We emphasize that the right unit sections associated with spin frames are
\textit{not} constant in any covariant way. In fact, we have the following.

\begin{proposition}
\label{cov der of 1^r}Let $\mathbf{1}_{\Xi}^{r}\in\sec\mathcal{C}%
\ell_{\mathrm{Spin}_{1,3}^{e}}^{r}(M)$ be the right unit section associated to
the spin frame $\Xi$. Then%
\begin{equation}
\mathbf{\nabla}_{e_{a}}^{s}\mathbf{1}_{\Xi}^{r}=-{\frac{1}{2}}\mathbf{1}_{\Xi
}^{r}\omega_{e_{a}}\text{,} \label{deriv of 1}%
\end{equation}
where $\omega_{e_{a}}$ is the connection 1-form (proposition \ref{DERCLIFFORD}%
)\ written in the basis $\{e_{a}\}$.
\end{proposition}

\textbf{Proof.} It follows from Eq.~(\ref{NCS000}) of the
Appendix.$\blacksquare$

\subsection{Representatives of \emph{DHSF} on the Clifford Bundle}

Let $\{\mathbf{E}^{a}\}$ be, as before, a fixed orthonormal basis of
$\mathbb{R}^{1,3}\subseteq\mathbb{R}_{1,3}.$ Remember that these objects are
fundamental to the Dirac equation (\ref{DE for Psi in Cl_left}) in terms of
sections $\mathbf{\Psi}$ of $\mathcal{C}\ell_{\mathrm{Spin}_{1,3}^{e}}%
^{l}(M,g)$:%
\[
D^{s}\mathbf{\Psi E}^{21}-qA\mathbf{\Psi}=m\mathbf{\Psi E}^{0}.
\]
Let $\Xi\in\sec P_{\mathrm{Spin}_{1,3}^{e}}(M)$ be a spin frame on $M\ $and
define the sections $\mathbf{1}_{\Xi}^{l},$ $\mathbf{1}_{\Xi}^{r}$ and $e_{a}%
$, respectively on $\mathcal{C}\ell_{\mathrm{Spin}_{1,3}^{e}}^{l}(M),$
$\mathcal{C}\ell_{\mathrm{Spin}_{1,3}^{e}}^{r}(M)$ and $\mathcal{C}\ell(M,g)$,
as above. Now we can use proposition \ref{E_a in terms of e_a} to write the
above equation in terms of sections of $\mathcal{C}\ell(M,g):$%
\begin{equation}
(D^{s}\mathbf{\Psi)1}_{\Xi}^{r}e^{21}\mathbf{1}_{\Xi}^{l}-qA\mathbf{\Psi
}=m\mathbf{\Psi1}_{\Xi}^{r}e^{0}\mathbf{1}_{\Xi}^{l}.
\end{equation}
Right-multiplying by $\mathbf{1}_{\Xi}^{r}$ yields, using proposition
\ref{E_a in terms of e_a},%
\begin{equation}
e^{a}(\mathbf{\nabla}_{e_{a}}^{s}\mathbf{\Psi)1}_{\Xi}^{r}e^{21}%
-qA\mathbf{\Psi1}_{\Xi}^{r}=m\mathbf{\Psi1}_{\Xi}^{r}e^{0}.
\end{equation}

It follows from proposition \ref{Leib 1} that
\begin{align}
(\mathbf{\nabla}_{e_{a}}^{s}\mathbf{\Psi)1}_{\Xi}^{r}  &  =\mathbf{\nabla
}_{e_{a}}(\mathbf{\Psi1}_{\Xi}^{r})-\mathbf{\Psi\nabla}_{e_{a}}^{s}%
(\mathbf{1}_{\Xi}^{r})\nonumber\\
&  =\mathbf{\nabla}_{e_{a}}(\mathbf{\Psi1}_{\Xi}^{r})+{\frac{1}{2}%
}\mathbf{\Psi1}_{\Xi}^{r}\omega_{a}, \label{Leib 3}%
\end{align}
where proposition \ref{cov der of 1^r} was employed in the last step.
Therefore%
\begin{equation}
e^{a}\left[  \mathbf{\nabla}_{e_{a}}(\mathbf{\Psi1}_{\Xi}^{r})+{\frac{1}{2}%
}\mathbf{\Psi1}_{\Xi}^{r}\omega_{a}\right]  e^{21}-qA(\mathbf{\Psi1}_{\Xi}%
^{r})=m(\mathbf{\Psi1}_{\Xi}^{r})e^{0}.
\end{equation}
Thus it is natural to define, for each spin frame $\Xi$, the Clifford field
$\psi_{\Xi}\in\sec C\ell(M,g)$ (see proposition
\ref{sec(Cl_left) x sec(Cl_right)}) by%
\begin{equation}
\mathit{\psi}_{\Xi}:=\mathbf{\Psi1}_{\Xi}^{r}. \label{definition of phi_Xi}%
\end{equation}
We then have%
\begin{equation}
e^{a}\left[  \mathbf{\nabla}_{e_{a}}\mathit{\psi}_{\Xi}+{\frac{1}{2}%
}\mathit{\psi}_{\Xi}\omega_{a}\right]  e^{21}-qA\mathit{\psi}_{\Xi
}=m\mathit{\psi}_{\Xi}e^{0}. \label{EDHprov}%
\end{equation}

A comment about the nature of spinors is in order. As we repeatedly\textit{
}said in the previous sections, spinor fields should not be ultimately
regarded as fields of multivectors (or multiforms), for their behavior under
Lorentz transformations is not tensorial (they are able to distinguish between
$2\pi$ and $4\pi$ rotations). So, how can the identification above be correct?
The answer is that the definition in Eq.~(\ref{definition of phi_Xi}) is
intrinsically spin-frame dependent. Clearly, this is the price one ought to
pay if one wants to make sense of the procedure of representing spinors by
differential forms.

Note also that the covariant derivative acting on $\mathit{\psi}_{\Xi}$ in
Eq.~(\ref{EDHprov}) is the tensorial covariant derivative $\mathbf{\nabla}%
_{V}$ on $C\ell(M,g)$, as it should be. However, we see from the expression
above that $\mathbf{\nabla}_{V}$ always acts on $\mathit{\psi}_{\Xi}$ together
with the term ${\frac{1}{2}}\mathit{\psi}_{\Xi}\omega_{a}$. Therefore, it is
natural to define an \textquotedblleft effective
covariant\ derivative\textquotedblright\ $\mathbf{\nabla}_{V}^{(s)}$ acting on
$\mathit{\psi}_{\Xi}$ by%
\begin{equation}
\mathbf{\nabla}_{e_{a}}^{(s)}\mathit{\psi}_{\Xi}:=\mathbf{\nabla}%
_{a}\mathit{\psi}_{\Xi}+{\frac{1}{2}}\mathit{\psi}_{\Xi}\omega_{a}%
.\label{hahaha}%
\end{equation}
Then, proposition \ref{DERCLIFFORD} yields
\begin{equation}
\mathbf{\nabla}_{e_{a}}^{(s)}\mathit{\psi}_{\Xi}=\partial_{e_{a}}%
(\mathit{\psi}_{\Xi})+{\frac{1}{2}}\omega_{a}\mathit{\psi}_{\Xi}%
\text{,}\label{hahaha'}%
\end{equation}
which emulates the spinorial covariant derivative\footnote{This is the
derivative used in \cite{50}, there introduced in an \emph{ad hoc} way.}, as it
should. We observe moreover that if $U\in\sec C\ell(M,g)$ and if
$\mathit{\psi}_{\Xi}\in\sec C\ell(M,g)$ is a representative of a
Dirac-Hestenes spinor field then%
\begin{equation}
\mathbf{\nabla}_{e_{a}}^{(s)}\left(  U\mathit{\psi}_{\Xi}\right)  =\left(
\mathbf{\nabla}_{e_{a}}U\right)  \mathit{\psi}_{\Xi}+U\mathbf{\nabla}_{e_{a}%
}^{(s)}\mathit{\psi}_{\Xi}\label{HAHAHA}%
\end{equation}

With this notation, we finally have the Dirac-Hestenes equation for the
\emph{representative} Clifford field $\mathit{\psi}_{\Xi}\in\sec C\ell(M,g)$,
on a Lorentzian spacetime\footnote{The \textit{DHE} on a Riemann-Cartan
spacetime will be discussed in another publication.}:%
\begin{equation}
e^{a}\mathbf{\nabla}_{e_{a}}^{(s)}\mathit{\psi}_{\Xi}e^{21}-qA\mathit{\psi
}_{\Xi}=m\mathit{\psi}_{\Xi}e^{0}, \label{DHE in a RC spacetime}%
\end{equation}
where $\mathit{\psi}_{\Xi}$ is the representative of a \textit{DHSF}
$\mathbf{\Psi}$ of $\mathcal{C}\ell_{\mathrm{Spin}_{1,3}^{e}}^{l}(M,g)$,
relative to the spin frame $\Xi$.

Let us finally show that this formulation recovers the usual transformation
properties characteristic of the Hestenes's formalism as described, e.g., in
\cite{50}. For that matter, consider two spin frames $\Xi,\Xi^{\prime}\in\sec
P_{\mathrm{Spin}_{1,3}^{e}}(M),$ with $\Xi^{\prime}(x)=\Xi(x)u(x)$, where
$u(x)\in\mathrm{Spin}_{1,3}^{e}$. It follows from proposition
\ref{transformation of 1_right} that $\mathit{\psi}_{\Xi^{\prime}%
}=\mathbf{\Psi1}_{\Xi^{\prime}}^{r}=\mathbf{\Psi}u^{-1}\mathbf{1}_{\Xi}%
^{r}=\mathbf{\Psi1}_{\Xi}^{r}U^{-1}=\mathit{\psi}_{\Xi}U^{-1}.$ Therefore, the
various spin frame dependent Clifford fields from Eq.
(\ref{DHE in a RC spacetime}) transform as%
\begin{align}
e_{a}' &  =Ue_{a}U^{-1},\label{transf dhsf}\\
\mathit{\psi}_{\Xi^{\prime}} &  =\mathit{\psi}_{\Xi}U^{-1}.\nonumber
\end{align}
These are exactly the transformation rules one expects from fields satisfying
the Dirac-Hestenes equation (see, e.g., \cite{50}).

\subsection{Bilinear Covariants}

\subsubsection{ Bilinear Covariants Associated to a \textit{DHSF}}

We are now\ in position to give a precise definition of the bilinear
covariants of the Dirac theory, associated with a given \textit{DHSF.}

\begin{definition}
Recalling that $\bigwedge\nolimits^{p}(M)\hookrightarrow\mathcal{C}\ell(M,g)$,
$p=0,1,2,3,4$, and recalling propositions \ref{sec(Cl_left) x sec(Cl_right)}
and \ref{sec(CL_right) on sec(Cl_left)}, the bilinear covariants associated to
a DHSF $\mathbf{\Psi}\in\sec C\ell_{\mathrm{Spin}_{1,3}^{e}}^{l}(M)$ (and
$\tilde{\mathbf{\Psi}}\in\sec\mathcal{C}\ell_{\mathrm{Spin}_{1,3}^{e}}%
^{r}(M))$\ are the following sections of $\mathcal{C}\ell(M,g)$:%
\begin{align}
S &  =\mathbf{\Psi}\tilde{\mathbf{\Psi}}=\sigma+e_{5}\,\omega\in\sec(\bigwedge
\nolimits^{0}(M)+\bigwedge\nolimits^{4}(M)),\label{fierz}\\
J &  =\mathbf{\Psi E}_{0}\tilde{\mathbf{\Psi}}\in\sec\bigwedge\nolimits^{1}%
(M),\text{ }K=\mathbf{\Psi\mathbf{E}}_{3}\tilde{\mathbf{\Psi}}\in\sec\bigwedge
\nolimits^{1}(M),\nonumber\\
M &  =\mathbf{\Psi\mathbf{E}}_{12}\tilde{\mathbf{\Psi}}\in\sec\bigwedge\nolimits^{2}%
(M),\nonumber
\end{align}
where $\mathbf{\Psi=\Psi}\frac{1}{2}(1+\mathbf{E}_{0}),$ and $e_{5}=e_{0}%
e_{1}e_{2}e_{3}$.
\end{definition}

\begin{remark}
Of course, since all bilinear covariants in Eq.~(\textit{\ref{fierz}}) are
sections of $\mathcal{C}\ell(M,g)$, they have the right transformation
properties under arbitrary local Lorentz transformations, as required. As
shown, e.g., in \cite{39} \ these bilinear covariants and their Hodge duals
satisfy a set of identities, called the Fierz identities (see, e.g., \cite{50})
that are crucial for the physical interpretation of the Dirac equation (in
first and second quantizations).
\end{remark}

\begin{remark}
Crumeyrolle \cite{14} gives the name of \emph{amorphous} spinor fields to
ideal sections of the Clifford bundle $\mathcal{C}\ell(M,g)$. Thus an
amorphous spinor field $\mathbf{\phi}$ is a section of $\mathcal{C}\ell(M,g)$
such that $\mathbf{\phi}\mathrm{P}=\mathbf{\phi}$, where $\mathrm{P=P}^{2}$ is
an idempotent section of $\mathcal{C}\ell(M,g)$. However, these fields and
also the so-called Dirac-K\"{a}hler fields (\cite{23},\cite{24K}), which are
also sections of $\mathcal{C}\ell(M,g)$, cannot be used in a physical theory
of fermion fields since they do not have the correct transformation law under
a Lorentz rotation of the local \textit{spin frame}.
\end{remark}

\subsubsection{ Bilinear Covariants Associated with a representative of a
\textit{DHSF}}

We note that the bilinear covariants, when written in terms of $\mathit{\psi
}_{\Xi}:=\mathbf{\Psi1}_{\Xi}^{r}$, read (from proposition
\ref{E_a in terms of e_a}) as%
\begin{align*}
S &  =\mathit{\psi}_{\Xi}\mathit{\tilde{\psi}}_{\Xi}=\sigma+e_{5}\,\omega
\in\sec(\bigwedge\nolimits^{0}(M)+\bigwedge\nolimits^{4}(M)),\\
J &  =\mathit{\psi}_{\Xi}e_{0}\mathit{\tilde{\psi}}_{\Xi}\in\sec
\bigwedge\nolimits^{1}(M),\text{ }K=\mathit{\psi}_{\Xi}e_{3}\mathit{\tilde
{\psi}}_{\Xi}\in\sec\bigwedge\nolimits^{1}(M),\\
M &  =\mathit{\psi}_{\Xi}e_{1}e_{2}\mathit{\tilde{\psi}}_{\Xi}\mathbf{\in}%
\sec\bigwedge\nolimits^{2}(M),
\end{align*}
where $e_{5}=e_{0}e_{1}e_{2}e_{3}$. These are all intrinsic quantities, as
they should be.

\subsection{Electromagnetic Gauge Invariance of the \emph{DHE}}

\begin{proposition}
The DHE is invariant under electromagnetic gauge transformations%
\begin{align}
\mathit{\psi}_{\Xi} &  \mapsto\mathit{\psi}_{\Xi}^{\prime}=\psi_{\Xi
}e^{qe_{21}\chi},\label{GL1a}\\
A &  \mapsto A+\partial\chi\text{, }\label{GL1b}\\
\omega_{e_{a}} &  \mapsto\omega_{e_{a}}%
\end{align}
where $\mathit{\psi}_{\Xi},\mathit{\psi}_{\Xi}^{\prime}\in\sec\mathcal{C}%
\ell^{0}(M,g)$, $A\in\sec\bigwedge\nolimits^{1}(M)\subset\sec C\ell(M,g)$ and
where $\chi\in\sec\bigwedge\nolimits^{0}(M)\subset\sec C\ell(M,g)$ is a gauge function.
\end{proposition}

\textbf{Proof. }It is a direct calculation.$\blacksquare$

But, what are the meanings of these transformations? Eq.(\ref{GL1a}) looks
similar to Eq.~(\ref{transf dhsf}) defining the change of a representative of
a \emph{DHSF} once we change spin frame, but here we have an active
transformation, since we did \textit{not }change the spin frame. On the other
hand, Eq.~(\ref{GL1b}) does not correspond either to a passive (no
transformation at all) or active local Lorentz transformation for $A$.
Nevertheless, writing $\chi=\theta/2$ yields
\begin{align}
e^{-qe^{21}\theta/2}e^{0}e^{qe^{21}\theta/2} &  =e^{\prime0}=e^{0}%
\text{,}\nonumber\\
e^{-qe^{21}\theta/2}e^{1}e^{qe^{21}\theta/2} &  =e^{\prime1}=\cos
q\theta~e^{1}+\sin q\theta~e^{2},\nonumber\\
e^{-qe^{21}\theta/2}e^{2}e^{qe^{21}\theta/2} &  =e^{\prime2}=-\sin
q\theta~e^{1}+\cos q\theta~e^{2},\nonumber\\
e^{-qe^{21}\theta/2}e^{3}e^{qe^{21}\theta/2} &  =e^{\prime3}=e^{3}.\label{GI4}%
\end{align}
We see that Eqs.~(\ref{GI4}) define a spin frame $\Xi^{\prime}$ to which
corresponds, as we already know, a basis $\ \{e^{\prime0},e^{\prime
1},e^{\prime2},e^{\prime3}\}$ for $\bigwedge\nolimits^{1}(M)\hookrightarrow
C\ell(M,g)$. We can then think of the electromagnetic gauge transformation as
a rotation in the spin plane $e^{21}$ by identifying $\psi_{\Xi}^{\prime}$ in
Eq.(\ref{GL1a}) with $\psi_{\Xi^{\prime}}$, the representative of the
\emph{DHSF} in the spin frame $\Xi^{\prime}$ and by supposing that instead of
transforming the spin connection $\omega_{e_{a}}$ as in
Eq.~(\ref{connection transf}) it is taken as fixed and instead of maintaining
the electromagnetic potential $A$ fixed it is transformed as in
Eq.~(\ref{GL1b}). We observe that, since in the theory of the gravitational
field $\omega_{e_{a}}$ is associated with some aspects of that field, our
interpretation for the electromagnetic gauge transformation suggests a
possible nontrivial coupling between electromagnetism and gravitation,
if the Dirac-Hestenes equation is taken as a fundamental
representation of fermionic matter. We will explore this possibility in
another publication.

\section{Conclusions}

In this paper, we hope to have clarified the ontology of Dirac-Hestenes spinor
fields (on a general spacetime $\mathfrak{M=}(M,g,\nabla,\tau_{g},\uparrow)$
of the Riemann-Cartan type admitting a spin structure) and its relationship
with sums of even multivector fields (or differential forms). This has been
achieved through the introduction of the Clifford bundle of multivector fields
($\mathcal{C\ell}(M,g)$) and the \emph{left (}$\mathcal{C\ell}_{\mathrm{Spin}%
_{1,3}^{e}}^{l}(M)$\emph{) }and \emph{right} ($\mathcal{C\ell}_{\mathrm{Spin}%
_{1,3}^{e}}^{r}(M)$) spin-Clifford bundles on a spin manifold $(M,g)$, as well
as a study of the relations among these bundles. Left algebraic spinor fields
and Dirac-Hestenes spinor fields (both fields are sections of $\mathcal{C\ell
}_{\mathrm{Spin}_{1,3}^{e}}^{l}(M)$) have been defined and the relation
between them has been established. Moreover, a consistent Dirac equation for
a \textit{DHSF }$\mathbf{\Psi}\in\sec\mathcal{C\ell}_{\mathrm{Spin}_{1,3}^{e}%
}^{l}(M)$ (denoted \textit{DE}$\mathcal{C\ell}^{l})$ on a Lorentzian spacetime
was found. We succeeded also in obtaining in a consistent way a
\textit{representation} of the \textit{DE}$\mathcal{C\ell}^{l}$ in the
Clifford bundle. It is such equation satisfied \ by Clifford fields
$\mathit{\psi}_{\Xi}\in\sec\mathcal{C\ell}(M,g)$ that we called the
Dirac-Hestenes equation (\textit{DHE}). This means that to each \textit{DHSF
}$\mathbf{\Psi}\in\sec$ $\mathcal{C\ell}_{\mathrm{Spin}_{1,3}^{e}}^{l}(M)$ and
to each spin frame $\Xi\in\sec P_{\mathrm{Spin}_{1,3}^{e}}(M)$ there is a
well-defined even nonhomogeneous multivector field
$\mathit{\psi}_{\Xi}\in\sec\mathcal{C\ell}(M,g)$ (\textit{E}\emph{MFS})
associated with $\Psi$. Such a \emph{EMFS} is called a \textit{representative}
of the \emph{DHSF} on the given spin frame. And, of course, such a
\emph{EMFS} (the representative of the \emph{DHSF}) is \emph{not} a spinor field.
With this crucial distinction between a \emph{DHSF} and their \emph{EMFS}
representatives we presented a consistent theory for Clifford and spinor
fields of all kinds.

We emphasize that the \textit{DE}$\mathcal{C\ell}^{l}$ and the \textit{DHE},
although related, are of different mathematical natures. This issue has been
\ particularly scrutinized in sections 4 and 5. We studied also the local
Lorentz invariance and the electromagnetic gauge invariance and showed that
only for the \textit{DHE }such transformations are of the same mathematical
nature, something that suggests by itself a possible link between them.

\bigskip

\textbf{Acknowledgments}: The authors are grateful to Doctors V. V.
Fern\'{a}ndez, A. Jadczyk, P. Lounesto, D. Miralles, A. M. Moya, I. Porteous
and J. Vaz, Jr. for discussions and to F. G. Rodrigues for help with latex.
They are also grateful to an anonymous referee for important observations.
W.A.R. is also grateful to CNPq for a research grant (contract 201560/82-8) and
to the Department of Mathematical Sciences of the University of Liverpool for
providing a Visiting Professor position and an enjoyable atmosphere for
research during the academic year 2001/2002. R.A.M. is grateful to FAPESP
(process number 98/16486-8). This paper is dedicated to the late P. Lounesto.

\appendix

\section{ Covariant Derivatives of Clifford and Spinor Fields}

\subsection{Covariant Derivative of Clifford Fields}

In this appendix, $(M,g,\mathbf{\nabla,\tau}_{g},\uparrow)$ denotes a general
\emph{Riemann-Cartan} spacetime (definition 3). Since $\mathcal{C}%
\ell(M,g)=\tau M/J(M,g)$, it is clear that any metric compatible
($\mathbf{\nabla}g=0$) connection defined in $\tau M$ passes to the quotient
$\tau M/J(M,g)$, and thus defines an algebra bundle connection \cite{14}. In
this way, the covariant derivative of a Clifford field $A\in\sec
\mathcal{C}\ell(M,g)$ is completely determined.

We will find \ formulas for the covariant derivative of Clifford fields and of
\emph{DHSF using }the general theory of connections on principal bundles and
covariant derivatives on associated vector bundles, as described in many
excellent textbooks, e.g., (\cite{12},\cite{20},\cite{42},\cite{43}).

Let $(E,M,\mathbf{\pi}_{1},G,F)$, denoted by $E=P\times_{\rho}F$, be a vector
bundle associated to a PFB $(P,M,\mathbf{\pi},G)$ by the linear representation
$\rho$ of $G$ in $F=\mathbf{V}$.

\begin{definition}
Let $\sigma:{}\mathbb{R}\supset I\rightarrow M,$ $t\mapsto\sigma(t)$ be a
curve in $M$ with $x_{0}=\sigma(0)\in M$, and let $p_{0}\in\mathbf{\pi}%
^{-1}(x_{0})$. The parallel transport of $p_{0}$ along $\sigma$ is given by
the curve $\hat{\sigma}:{}\mathbb{R}\supset I\rightarrow P,t\mapsto\hat
{\sigma}(t)$ defined by
\begin{equation}
\frac{d}{dt}\hat{\sigma}(t)=\mathbf{\Gamma}_{p}\left( \frac{d}{dt}\sigma(t) \right),
\label{4.13}
\end{equation}
with $p_{0}=\hat{\sigma}(0)$ and $\mathbf{\pi(}\hat{\sigma}(t))=\sigma(t)$. We
also denote $p_{\parallel t}=\hat{\sigma}(t)$.
\end{definition}

In Eq.~(\ref{4.13}), $\mathbf{\Gamma}_{p}:T_{x}M\rightarrow T_{p}P$
is a connection\footnote{See, e.g., definition (a) on page 358 of
\cite{12}.} on $(P,M,\mathbf{\pi},G)$.

Consider the trivializations of $P$
\[
\Phi_{i}:\mathbf{\pi}^{-1}(U_{i})\rightarrow U_{i}\times G,\quad\Phi
_{i}(p)=(\pi(p),\phi_{i,x}(p)),
\]
and $E$%
\[
\Xi_{i}:\mathbf{\pi}_{1}^{-1}(U_{i})\rightarrow U_{i}\times F,\quad\Xi
_{i}(q)=(\mathbf{\pi}_{1}(q),\chi_{i}(q))=\left(  x,\chi_{i}(q)\right).
\]
Then, we have the following.

\begin{definition}
\label{def transp para}The \emph{parallel transport} of
$\mathbf{\Psi}_{0}\in E$, $\mathbf{\pi}_{1}(\mathbf{\Psi}_{0})=x_{0}$,
along the curve $\sigma:\mathbb{R}\supset I\rightarrow M$, $t\mapsto
\sigma(t)$, from $x_{0}=\sigma(0)\in M$ to $x=\sigma(t)$ is the element
$\mathbf{\Psi}_{\mathbf{\parallel}t}\in E$ such that:
\end{definition}

(i) $\mathbf{\pi}_{1}(\mathbf{\Psi}_{\mathbf{\parallel}t})=x$;

(ii) $\chi_{i}(\mathbf{\Psi}_{\mathbf{\parallel}t})=\rho(\phi_{i}(p_{\parallel
t})\circ\phi_{i}(p_{0})^{-1})\chi_{i}(\mathbf{\Psi}_{0})$;

(iii) $p_{\parallel t}\in P$ is the parallel transport of $p_{0}\in P$ along
$\sigma$ from $x_{0}$ to $x$.

\begin{definition}
\noindent Let $v$ be a vector at $x_{0}$ tangent to the curve $\sigma$ (as
defined above). The covariant derivative of $\mathbf{\Psi\in}\sec E$ along $v$
is denoted $(D_{v}^{E}\mathbf{\Psi})_{x_{0}}\mathbf{\in}\sec E$ and
\begin{equation}
(D_{v}^{E}\mathbf{\Psi})(x_{0})\equiv(D_{v}^{E}\mathbf{\Psi})_{x_{0}}%
=\lim_{t\rightarrow0}\frac{1}{t}(\mathbf{\Psi}_{\mathbf{\parallel}t}%
^{0}-\mathbf{\Psi}_{0}), \label{4.66}%
\end{equation}
where $\mathbf{\Psi}_{\mathbf{\parallel}t}^{0}$ is the parallel transport of
the vector\ $\mathbf{\Psi}_{t}\equiv\mathbf{\Psi(}\sigma\mathbf{(}%
t\mathbf{))}$ of the given section $\mathbf{\Psi\in}\sec E$ along $\sigma$
from $\sigma(t)$ to $x_{0}$. The only requirements on $\sigma$ are that
$\sigma(0)=x_{0}$ and
\begin{equation}
\left.  \frac{d}{dt}\sigma(t)\right\vert _{t=0}=v. \label{4.67}%
\end{equation}

\end{definition}

\begin{proposition}
\label{DERCLIFFORD}Let $V\in\sec TM$. The covariant derivative of a Clifford
field $A\in\sec\mathcal{C}\ell(M,g)$ is given by
\begin{equation}
\mathbf{\nabla}_{V}A=V(A)+{\frac{1}{2}[}\omega_{V},A],\label{N4.00}%
\end{equation}
where $V(A):=V(A^{I})e_{I}$ and $\omega_{V}$ is the connection 1-form
$V\mapsto\omega_{V}=-\frac{1}{2}V^{a}\Gamma_{abc}e^{b}\wedge e^{c}$, written
in the basis $\{e_{a}\}$, with $\Gamma_{abc}$ given by $\mathbf{\nabla}%
_{e_{a}}e_{b}=\Gamma_{ab}{}^{c}e_{c}=\Gamma_{abc}e^{c}$.
\end{proposition}

\textbf{Proof \ }Writing $A(t)=A(\sigma(t))$ in terms of the multivector basis
$\{e_{I}\}$ of sections associated to a given spin frame, as in section
\ref{sec fid}, we have $A(t)=A^{I}(t)e_{I}(t)=A^{I}(t)[(\Xi(t),E_{I}%
)]=[(\Xi(t),A^{I}(t)E_{I})]=[(\Xi(t),a(t))],$ with $a(t):=A^{I}(t)E_{I}%
\in\mathbb{R}_{1,3}$. If follows from item (ii) of definition
\ref{def transp para} that%
\begin{equation}
A_{||t}^{0}=[(\Xi(0),g(t)a(t)g(t)^{-1})]\label{A^0_||t}%
\end{equation}
for some $g(t)\in\mathrm{Spin}_{1,3}^{e}$, with $g(0)=1$. Thus%
\begin{align*}
\lim_{t\rightarrow0}\frac{1}{t}(g(t)a(t)g(t)^{-1}-a(0)) &  =\left[  \frac
{dg}{dt}ag^{-1}+g\frac{da}{dt}g^{-1}+ga\frac{dg^{-1}}{dt}\right]  _{t=0}=\\
&  =\dot{a}(0)+\dot{g}(0)a(0)-a(0)\dot{g}(0)=\\
&  =V(A^{I})E_{I}+[\dot{g}(0),a(0)],
\end{align*}
where $\dot{g}(0)\in\operatorname{Lie}(\mathrm{Spin}_{1,3}^{e})=\Lambda
^{2}(\mathbb{R}^{1,3})$. Therefore%
\[
\nabla_{V}A=V(A^{I})e_{I}+\frac{1}{2}[\omega_{V},A],
\]
for some $\omega_{V}\in\sec\bigwedge\nolimits^{2}(M).$ In particular,
calculating the covariant derivative of the basis 1-vector fields $e_{b}$
yields $V^{a}\Gamma_{ab}{}^{c}e_{c}=\mathbf{\nabla}_{V}e_{b}=\frac{1}%
{2}[\omega_{V},e_{b}]$, so that $\omega_{V}=-\frac{1}{2}V^{a}\Gamma_{abc}%
e^{b}\wedge e^{c}$.$\blacksquare$

\begin{remark}
Equation (\ref{N4.00}) shows that the covariant derivative preserves the
degree of a homogeneous Clifford field, as can be easily verified.
\end{remark}

The general formula Eq.~(\ref{N4.00}) and the associative law in the Clifford
algebra immediately yields the following.

\begin{corollary}
The covariant derivative $\mathbf{\nabla}_{V}$ on $\mathcal{C}\ell(M,g)$ acts
as a derivation on the algebra of sections, i.e., for $A,B\in\sec
\mathcal{C}\ell(M,g)$ and $V\in\sec TM$, it holds that
\begin{equation}
\mathbf{\nabla}_{V}(AB)=(\mathbf{\nabla}_{V}A)B+A(\mathbf{\nabla}%
_{V}B)\label{Leibnitz-Clifford}%
\end{equation}

\end{corollary}

Under a change of gauge (local Lorentz transformation) $e^{a}\mapsto e^{\prime
a}=Ue^{a}U^{-1}$, with $U\in\sec\mathcal{C}\ell(M,g),U\tilde{U}=\tilde{U}U=1$,
the corresponding transformation law for $\omega_{V}$ is as follows.

\begin{corollary}
Under a change of gauge (local Lorentz transformation), $\omega_{V}$
transforms as
\begin{equation}
\frac{1}{2}\omega_{V}\mapsto U\frac{1}{2}\omega_{V}U^{-1}+(\mathbf{\nabla}%
_{V}U)U^{-1}, \label{connection transf}%
\end{equation}

\end{corollary}

\textbf{Proof. }It is a simple calculation using Eq.(\ref{N4.00}%
).$\blacksquare$

\subsection{ Covariant Derivatives of Spinor Fields}

The spinor bundles introduced in section 2, like $I(M)=P_{\mathrm{Spin}%
_{1,3}^{e}}(M)\times_{l}I$, $I=\mathbb{R}_{1,3}\frac{1}{2}(1+E_{0})$,
$\mathcal{C}\ell_{\mathrm{Spin}_{1,3}^{e}}^{l}(M)$, and $\mathcal{C}%
\ell_{\mathrm{Spin}_{1,3}^{e}}^{r}(M)$ (and subbundles) are examples of vector
bundles. Thus, the general theory of covariant derivative operators on
associated vector bundles can be used (as in the previous section) to obtain
formulas for the covariant derivatives of sections of these bundles. Given
$\Psi\in\sec\mathcal{C}\ell_{\mathrm{Spin}_{1,3}^{e}}^{l}(M)$ and $\Phi\in
\sec\mathcal{C}\ell_{\mathrm{Spin}_{1,3}^{e}}^{r}(M)$, we denote the
corresponding covariant derivatives by $\mathbf{\nabla}_{V}^{s}\Psi$ and
$\mathbf{\nabla}_{V}^{s}\Phi$ \footnote{Recall that $I^{l}(M)\hookrightarrow
C\ell_{\mathrm{Spin}_{1,3}^{e}}^{l}(M)$ and $I^{r}(M)\hookrightarrow
C\ell_{\mathrm{Spin}_{1,3}^{e}}^{r}(M)$.}.

\begin{proposition}
Given $\Psi\in\sec\mathcal{C}\ell_{\mathrm{Spin}_{1,3}^{e}}^{l}(M)$ and
$\Phi\in\sec\mathcal{C}\ell_{\mathrm{Spin}_{1,3}^{e}}^{r}(M),$
\begin{align}
\mathbf{\nabla}_{V}^{s}\Psi &  =V(\Psi)+{\frac{1}{2}}\omega_{V}\Psi
,\label{NCS00}\\
\mathbf{\nabla}_{V}^{s}\Phi &  =V(\Psi)-{\frac{1}{2}}\Psi\omega_{V}.
\label{NCS000}%
\end{align}

\end{proposition}

\textbf{Proof. }It is analogous to that of proposition \ref{DERCLIFFORD}, with
the difference that Eq.~(\ref{A^0_||t}) should be substituted by $\Psi
_{||t}^{0}=[(\Xi(0),g(t)a(t))]$ and $\Phi_{||t}^{0}=[(\Xi(0),a(t)g(t)^{-1}%
)].\blacksquare$

\begin{proposition}
\label{Leib 1}Let $\mathbf{\nabla}$ be the connection on $\mathcal{C}%
\ell(M,g)$ to which $\mathbf{\nabla}^{s}$ is related. Then, for any $V\in\sec
TM$, $A\in\sec\mathcal{C}\ell(M,g)$, $\Psi\in\sec\mathcal{C}\ell
_{\mathrm{Spin}_{1,3}^{e}}^{l}(M)$) and $\Phi\in\sec\mathcal{C}\ell
_{\mathrm{Spin}_{1,3}^{e}}^{r}(M),$%
\begin{align}
\mathbf{\nabla}_{V}^{s}(A\Psi)  &  =A\mathbf{\nabla}_{V}^{s}\Psi
+(\mathbf{\nabla}_{V}A)\Psi,\label{SCD1}\\
\mathbf{\nabla}_{V}^{s}(\Phi A)  &  =\Phi\mathbf{\nabla}_{V}A\mathbf{+}%
(\mathbf{\nabla}_{V}^{s}\Phi)A. \label{SCD11}%
\end{align}

\end{proposition}

\textbf{Proof. }Recalling that $\mathcal{C}\ell_{\mathrm{Spin}_{1,3}^{e}}%
^{l}(M)$ ($\mathcal{C}\ell_{\mathrm{Spin}_{1,3}^{e}}^{r}(M)$) is a module over
$\mathcal{C}\ell(M,g),$ the result follows from a simple
computation.$\blacksquare$

Finally, let $\Psi\in\sec\mathcal{C}\ell_{\mathrm{Spin}_{1,3}^{e}}^{l}(M)$ be
such that $\Psi\mathbf{e}=\Psi$, where $\mathbf{e}^{2}=\mathbf{e}\in
\mathbb{R}_{1,3}$ is a primitive idempotent. Then, since
$\Psi\mathbf{e}=\Psi$,
\begin{align}
\mathbf{\nabla}_{V}^{s}\Psi &  \mathbf{=}\mathbf{\nabla}_{V}^{s}%
(\Psi\mathbf{e)}\mathbf{=}V(\Psi\mathbf{e})+\frac{1}{2}\omega_{V}%
\Psi\mathbf{e}\nonumber\\
&  =[V(\Psi)+\frac{1}{2}\omega_{V}\Psi]\mathbf{e=(\nabla}_{V}^{s}%
\Psi)\mathbf{e,} \label{SCD111}%
\end{align}
from where we verify that the covariant derivative of a \textit{LIASF} is
indeed a \textit{LIASF}.

\end{document}